\documentclass[fleqn,usenatbib]{mnras}

\usepackage[T1]{fontenc}
\usepackage{ae,aecompl}
\usepackage{color}
\usepackage{graphicx}	
\usepackage{amsmath}	
\usepackage{amssymb}	

\title{The fate of high-redshift massive compact galaxies}

\author[de la Rosa et al.]{%
Ignacio G. de la Rosa$^{1,2}$\thanks{E-mail: irosa@iac.es}, Francesco La Barbera$^3$, 
Ignacio Ferreras$^4$,
\newauthor
Jorge S\'anchez Almeida$^{1,2}$, Claudio Dalla Vecchia$^{1,2}$, 
Inma Mart\'inez-Valpuesta$^{1,2}$,
\newauthor
Martin Stringer$^{1,2}$
\\
$^1$ Instituto de Astrof\'\i sica de Canarias, C/ V\'\i a L\'actea
s/n, La Laguna, E-38205 La Laguna, Tenerife, Spain\\ 
$^2$ Departamento de Astrof\'\i sica,
Universidad de La Laguna, E-38206 La Laguna, Tenerife, Spain \\
$^3$ INAF - Osservatorio Astronomico di Capodimonte, Napoli, 80131, Italy \\
$^4$ Mullard Space Science Laboratory, University College London, 
Holmbury St Mary, Dorking, Surrey RH5 6NT
}

\date{Accepted 2016 January 13. Received 2015 December 24; in original form 2015 July 30}

\pubyear{2016}

\begin{document}
\label{firstpage}
\pagerange{\pageref{firstpage}--\pageref{lastpage}}
\maketitle

\begin{abstract}
Massive high-redshift quiescent compact galaxies (nicknamed {\it red nuggets}) have been traditionally connected to present-day elliptical galaxies, often overlooking the relationships that they may have with other galaxy types. We use large bulge-disk decomposition catalogues based on the Sloan Digital Sky Survey (SDSS) to check the hypothesis that red nuggets have {\it survived} as compact cores embedded inside the haloes or disks of present-day massive galaxies. In this study, we designate a {\it compact core} as the bulge component that satisfies a prescribed compactness criterion. Photometric and dynamic mass-size and mass-density relations are used to show that, in the inner regions of galaxies at $z\sim0.1$, there are {\it abundant} compact cores matching the peculiar properties of the red nuggets, an abundance comparable to that of red nuggets at $z\sim1.5$. Furthermore, the morphology distribution of the present-day galaxies hosting compact cores is used to demonstrate that, in addition to the standard channel connecting red nuggets with elliptical galaxies, a comparable fraction of red nuggets might have ended up embedded in disks. This result generalises the inside-out formation scenario; present-day massive galaxies can begin as dense spheroidal cores (red nuggets), around which either a spheroidal halo or a disk are formed later.
\end{abstract}




\section{Introduction}

An approximate census of massive galaxies at redshift $z\sim1.5$
provides a morphological classification into 40\% disks, 15\% extended
spheroids, 25\% compact spheroids and 20\% peculiar objects
\citep[e.g.][]{Peth15}. More than 9 Gyrs later, in the present-day
Universe, a similar census of massive galaxies gives 30\% disks, 65\%
extended spheroids and 5\% peculiar morphologies. Strikingly, compact
spheroids are virtually absent in the present-day Universe
\citep[e.g.][]{Tru09,Tay10,Saul15}. At $z\sim1.5$, the majority ($\sim$80 \%) of massive compact spheroids were already quiescent, while the
rest were actively forming stars. The massive high-redshift quiescent
compact galaxies have been termed {\it red nuggets}
\citep{Dam09}. Recent studies on the number density evolution of red
nuggets point to the star forming compact massive galaxies (sCMG) as
their probable progenitors \citep[e.g.][]{Barro13,vD15}. At redshift
$z\sim2.5$, the number density of sCMGs matched that of red nuggets.
By $z\sim1.5$, 1.5 Gyr later, the number density of red
nuggets has increased by an order of magnitude while that of sCMGs
has decreased by a factor of two. The straightforward
interpretation is that the red nuggets are the descendants of rapidly quenched sCMGs. But unlike this relative consensus on the origin of red nuggets, there is not the same agreement regarding their fate after $z\sim1.5$.

Red nuggets were discovered almost a decade ago
\citep[e.g.][]{Dad05,Tru06a,Lon07} and more than a hundred studies
have been carried out on their evolution. Since their discovery, red
nuggets have been linked to present-day elliptical galaxies; the two classes share a similar spheroidal morphology, and the fraction of elliptical galaxies (extended
spheroids) has increased by a factor of four in the last 9 Gyrs
\citep{Bui13}. But for the former population to have evolved into the latter, there must be a mechanism by which an ultra-compact spheroid with half-light radius
of $\sim$1 kpc can become a fully-fledged elliptical, three to six times
larger. Several physical processes, including major and minor mergers
\citep[e.g.][]{Naab09,Tru11} and puffing-up by AGN activity
\citep{Fan08}, have been proposed to explain the considerable growth
in size. The prevailing paradigm, put forward by \cite{Hop09} and
\cite{Bez09}, postulates an appealing scenario in which red nuggets
ended up in the centres of present-day giant ellipticals. This
proposal is consistent with the inside-out growth scenario
\citep[e.g.][]{LP03} in which compact cores are formed at high
redshift ($z\geqslant 2$) through highly dissipative processes
\citep[e.g.][]{Naab07,Dek09,Oser10,Joh12} and
then grow an extended stellar halo through dissipation-less minor
mergers \citep[e.g.][]{Ferr14,MorI16}.

In addition to this traditional evolutionary scenario, there is another potentially important channel leading from red nuggets to disk galaxy
bulges. The main observational evidences for
this scenario have been provided by \cite{Gra13}, \cite{DG13} and \cite{GraDS15}. \cite{Gra13} sample includes the set of 19 red nuggets
observed by \cite{Dam09} and a catalogue of few hundred bulges
compiled by \cite{GW08}. In two subsequent studies, \cite{DG13}
and \cite{GraDS15} carried out detailed bulge+disk (B+D) decompositions, but only for a small sample of present-day S0 galaxies, finding that S0 bulges are structurally similar to red nuggets.

Figure 1 of \cite{Gra13} presents an instructive sketch of mass-size and mass-density diagrams
for all the spheroidal families, from giant elliptical galaxies to
globular clusters. In the accompanying text, the author emphasises the
considerable overlap of the structural properties of red nuggets with
those of present-day bulges. Based on cosmological simulations, \cite{Zol15} and \cite{Tacch15} show that extended star-forming discs develop around red nuggets after compaction. 

The aim of the present study is to carry out a comprehensive analysis of the connection between red nuggets and present-day compact cores without any restriction on the morphology of the host
galaxies. Specifically, the hypothesis we want to test is that a significant fraction of red
nuggets have {\it survived}, masked inside spheroidal or
disc galaxies in the present-day Universe.

To support this hypothesis, we would need to find a sufficient population of compact cores embedded inside present-day galaxies, where:
\begin{itemize}
\item{{\it Core} is a generalization of the bulge component in B+D decomposition of galaxies with all morphologies.} 
\item{{\it Compact core} is a core that fulfills the same compactness criterion used for red nuggets. We test the robustness of our results using three different definition of compactness, from the literature.} 
\item{{\it Sufficient} means that the abundance of present-day red nugget-like cores should be approaching\footnote{Based on cosmological simulations \citep[]{Wellons15,Furlong15}, we might expect $\sim$ 20 \% of the red nuggets to have been consumed or destroyed in merger events, so the abundances should not necessarily match exactly.} that of the red nuggets at $z\sim1.5$.}
\end{itemize}
The present study takes advantage of SDSS-based catalogues containing B+D decomposition of hundreds of thousands of galaxies \citep[e.g.][]{Sim11,Meert15}. In particular, we have used the \cite{Men14} catalogue, itself a derivation of \cite{Sim11}, which provides stellar masses for the B and D structural components. We impose the B+D decomposition on the entire present-day galaxy sample, regardless of their morphology, designating as {\it cores} both the central components of ellipticals and the bulges of disc galaxies. After B+D decomposition, cores are treated as independent structures and evaluated for their compactness, according to standard criteria (see section 2).

The layout of the paper is the following: Section 2 is devoted to the
data description, including the selection of the z$\sim$0 compact
core sample, reference elliptical galaxies and the compilation of red nuggets from the literature. Section 3 is devoted to the results of the present study,
in particular the check of the working hypothesis, i.e. that red nuggets have survived as the compact cores of present-day galaxies. Discussion and conclusions follow in Sections 4 and 5. Decimal logarithms are used along the study. To keep consistency with other studies, we assume a $\Lambda$CDM cosmology with $\Omega_{\Lambda}$ = 0.7, $\Omega_{\rm M}$ = 0.3 and $H_0$ = 70 km s$^{-1}$ Mpc$^{-1}$.

\section{The Data}
\label{sec:data}

Several samples are extracted from the parent catalogue \citep{Men14}. First we introduce the compactness criteria, which allow the selection of structurally similar cores and red nuggets. Second, low-redshift samples, namely the compact cores and reference ellipticals, are introduced and third, the high-redshift red nugget sample is presented.

\subsection{Compactness criteria}
\label{sec:compact}

Our goal is to measure the abundance of z$\sim$0.1 galaxy cores structurally similar to red nuggets, and compare it to that of red nuggets at z$\sim$1.5. Therefore, information on the red nugget number density as a function of redshift becomes crucial for our work. Three different studies have addressed this issue in the literature, based on different criteria to define the red nuggets. As emphasised by \citet{Dam15}, the galaxy number densities and the structural parameters are rather sensitive to the choice of the compactness criterion, as well as to the stellar mass range probed. Consequently, we have to assure that each comparison is carried out under the same selection criteria. 

In the present study, circularised half-light radii $R_{\rm e}$ have been used, with
\begin{equation}
    \log (R_{\rm e}) = \log(R_{\rm e, a}) + 0.5 \log(b/a)
	\label{eq:radius}
\end{equation}
where $R_{\rm e, a}$ is the half-light radius along the major-axis and (b/a) is the axis ratio of the galaxy.  

In order to test the robustness of our results, we compute the number density of red nuggets using three alternative definitions of compactness:

$\bullet$ The criterion proposed by \citep[]{Barro13} (hereinafter Barro13), used by several authors \citep[e.g.][]{Barro15,Dam15,Pog13},
\begin{equation}
   \log (R_{\rm e}/{\rm kpc}) < (\log(M_{\star}/{\rm M}_\odot) - 10.3)/1.5~.
	\label{eq:barro13}
\end{equation}
This is generally associated with a mass limit $\log(M_{\star}/{\rm M}_\odot)\geq10.0$.

$\bullet$ A more restrictive compactness criterion, proposed by \cite{vD15} (hereinafter vDokkum15), 
\begin{equation}
     \log (R_{\rm e}/{\rm kpc}) < \log(M_{\star}/{\rm M}_\odot) - 10.7,
	\label{eq:vD15}
\end{equation}
with a $\log(M_\star/{\rm M}_\odot)\geq 10.6$ limit.

$\bullet$ Finally, the criterion proposed by \cite{vdW14} (hereinafter vdWel14)
\begin{equation}
    R_{\rm e, a}/(M_\star/10^{11}{\rm M}_\odot)^{0.75} < 2.5\,{\rm kpc},
	\label{eq:vdW14}
\end{equation}
with a $\log(M_\star/{\rm M}_\odot)\geq 10.7$ limit. It is worth mentioning that this criterion uses $R_{\rm e, a}$, the half-light radius measured along the galaxy major axis, instead of the more common circularized radius, $R_{\rm e}$. 

While we test our results on the abundances of red nuggets and compact cores using all three different sets of selection criteria (\S\,\ref{sec:numberdens}), for the first part of this paper (\S\S\,\ref{sec:photo} - \ref{sec:dyn}) we adopt the vDokkum15 definition. In fact, the definition of vdWel14 would not allow us to perform a direct comparison to other spheroidal systems (\S\,\ref{sec:photo}) as it is based on major-axis effective radii. On the other hand, the lower mass cutoff of Barro13 would imply a significantly higher contamination from star-forming galaxies in our sample of red nuggets, with respect to the vDokkum15 higher mass cutoff \citep[see][]{Mor13}.

\subsection{Low redshift samples}
\label{sec:lowz}
 
The parent sample used in our study is the \cite{Men14} catalogue
(hereafter Mendel14), an upgraded section of the \cite{Sim11} catalogue
(Simard11), which contains the B+D decomposition of more
than one million SDSS-DR7 galaxies. The main improvement to the original catalogue is
the stellar mass calculation of the bulge and
disk components. Stellar masses are computed through Spectral Energy Distribution (SED)-fitting to
the five SDSS-$ugriz$ bands, using dusty synthetic templates from
\cite{Con09} with a \citet{Chab03} stellar initial mass function (IMF). The Mendel14 catalogue
includes 657,996 galaxies from the SDSS-DR7 \citep{Aba09} with
spectroscopic redshifts and successful $ugriz$ B+D decomposition available. The
original 2D decomposition in Simard11 used three alternative galaxy
fitting models: 
\begin{enumerate}
\item{de Vaucouleurs bulge (fixed Sersic index $n_{\rm
  bulge}$ = 4) + exponential disk (Sersic index $n_{\rm disk}$ = 1),}
\item{exponential disk + free-$n_{\rm bulge}$, with $n_{\rm bulge}$
allowed to vary from 0.5 to 8 and} 
\item{a single component free
$n$-Sersic model to fit the whole galaxy ($n$-Sersic).}
\end{enumerate}
It is important to notice that Mendel14 stellar masses were computed using only two of
those fitting models: (ii) and (iii). The bias introduced by the use of (i)
instead of (ii) is discussed in Appendix~\ref{app:bias}. 

Two low-redshift subsamples are extracted from the  Mendel14 catalogue. The main one is the {\it compact core} subsample (\S\,\ref{sec:compl}), including galaxy cores that fulfill the vDokkum15 compactness criterion. The second subsample consists of bona-fide ellipticals (\S\,\ref{sec:ellipt}), and serves as a reference point for the study of galaxy structural properties (\S\,\ref{sec:photo}).

Bulge-disk decomposition is a delicate procedure demanding extra care to guarantee reliable outcomes. We have applied a rigorous screening, with eight steps, generally following the recommendations of the catalogue authors \citep[e.g.][]{Men14,Berg14,Bluck14}.

\subsubsection{Compact core Subsample Selection}
\label{sec:compl}

We select compact cores from the Mendel14 catalogue based on different criteria. This is summarized in Table~\ref{tab:subs}, where we also report the number of objects surviving each selection step.

First, we adopt a surface brightness limit of $\mu_{50,r} \le 23.0$ mag arcsec$^{-2}$ to assure a complete sample. Although the formal surface brightness limit of the SDSS spectroscopic sample is $\mu_{50,r} = 24.5$ mag arcsec$^{-2}$ \citep{Stra02}, the sample in the range $\mu_{50,r}$ 23.0 - 24.5 
mag arcsec$^{-2}$ is not complete because galaxies were targeted only under some particular local sky conditions. Second, we exclude objects with uncertain decomposition, using the parameter {\it dBD}, which is the offset between two alternative galaxy mass calculations, $M_B$ + $M_D$
(sum of independently derived bulge and disk masses) and $M_{B+D}$
(mass derived from the total bulge+disk photometry) in units of the
standard error. Significant inequality between these two masses is an indication of inconsistent decomposition, so systems with the cutoff $dBD > 1\sigma$, proposed by Mendel14, are excluded. These two cuts removed 9.5 \% of the objects, as seen in Table~\ref{tab:subs}

\begin{figure}
  \includegraphics[width=\columnwidth]{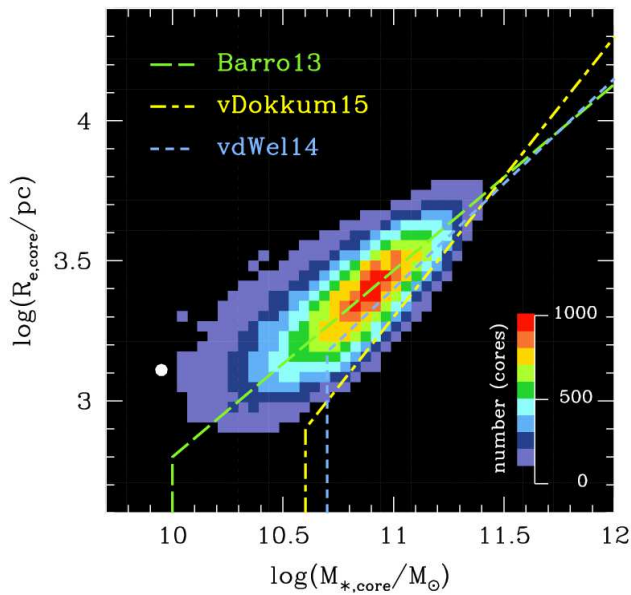}
  \caption{Bidimensional histogram showing the mass-size
    distribution for all the decomposed cores in our sample galaxies
    (203,369 objects, see Table~\ref{tab:subs}). Three compactness criteria, Barro13, vDokkum15 and vdWel14 (described in the text), are overplotted and compact
    cores are located below the dashed lines, rendering 65,637
    systems under the Barro13 criterion. The more restrictive vDokkum15 and vdWel14 criteria render 10,566 and 12,721 respectively. Note that the vdWel14 line is only approximate, because the criterion requires major-axis sizes instead of the circularized radii used in the plot. The isolated white dot marks the approximate location of the Milky Way's bulge (e.g. Di Matteo et al. 2015; Licquia \& Newman, 2015).}
  \label{fig:compactness}
\end{figure}

The third cutoff, limiting the redshift range, is set to achieve a
significant sample completeness in terms of the core stellar mass, $M_{\star,{\rm core}}$, the main reference parameter of this
study. The choice of redshift range, $0.025 \leq z \leq 0.15$, yields a significant completeness in our target mass range,
$\log(M_{\star,{\rm core}}/{\rm M}_\odot)\geq 10.0$, as discussed in Appendix~\ref{app:compl}. Notice that in the number density
estimations (\S\,\ref{sec:numberdens}), the V/Vmax statistic is
also used to correct for any residual incompleteness.

In the fourth step, we exclude galaxies reported to show anomalous B+D
decomposition, tagged `Type-4' in the Mendel14 classification. They comprise
galaxy/star superposition and inverted B+D profiles, in which a disk
dominates at the center and a bulge at the external parts. Following
\cite{Berg14}, fifth and sixth cutoffs were applied to promote
face-on inclinations and to avoid strong bars. Face-on galaxies (disk
inclination $\leq$ 60 degrees) are favoured in order to minimize
internal disk extinction on the bulge light, while high bulge
ellipticities are avoided to prevent strong bars being mistaken for
bulges (e$_{\rm core} <$ 0.6).

Finally, the potential impact of poorly resolved galaxy images on the
B+D decomposition is considered. We follow the study by \cite{Gad08},
which showed that the structural properties of cores can be reliably
retrieved provided that their effective radii are larger than $\sim$
80 per cent of the PSF half-width-at-half-maximum (HWHM), i.e. 
($R_{\rm e,core}$/PSF) $\geq$ 0.8. So far, by imposing these seven selection criteria, the parent sample has been reduced to 203,369 objects (31\% of the original sample; see Tab.~\ref{tab:subs}). The last selection is on the cores that fulfill the vDokkum15 compactness criterion (equation \ref{eq:vD15}). Additionally, Table \ref{tab:subs} shows the corresponding results for the other criteria (Equations \ref{eq:barro13} and \ref{eq:vdW14}). Figure \ref{fig:compactness} provides a visual display of the compact core selection criteria for both Barro13 and vDokkum15 on top of a 2D histogram including the 203,369 objects, which got through the first seven screening steps.  

The {\it compact core} definition is essential to the present
study and deserves further clarification. The B+D decomposition
is straightforward for disk galaxies, where a compact core is
simply a bulge that has satisfied the compactness criterion. A
decomposed elliptical galaxy should not be interpreted literally as
having physically meaningful bulge and a disk. In addition to the
central {\it core}, the external component can be a disk, a halo or
any significant departure from a single fit. Distinct inner components
have been extensively observed in ellipticals
\citep[e.g.][]{Cote06,Lau07}. More recently, \cite{Huang13} have shown
that more than 75 \% of their sample of 94 nearby elliptical galaxies
were better described when additional subcomponents were added to the
conventional single Sersic function.

\begin{table}
\caption{Subsample selection, listing the steps leading from the parent sample (Mendel14) to our {\it compact core} subsample. Steps are described in the main text.}
\label{tab:subs}
\centering
\begin{tabular}{ | l  | r | }
\hline
Filters & Objects \\
\hline 
Parent sample (Mendel14) & 657,996\\ 
Surface Brightness & 655,530 \\ 
Uncertain Decomposition & 595,402 \\ 
0.025 $\leq$ z $\leq$ 0.15 & 456,169 \\ 
Type $\neq$ 4 & 437,089 \\ 
${\rm Disk(i)\leq60^o}$ & 291,192 \\ 
${\rm e_{\rm core} < 0.6}$ & 249,990 \\ 
(R$_{\rm core}$/PSF) $\geq$ 0.8 & 203,369 \\
{\bf Compact cores (vDokkum15)} & {\bf 10,566} \\
\hline
Compact cores (Barro13) & 65,637 \\
Compact cores (vdWel14) & 12,721 \\
\hline
\end{tabular}
\end{table}

We notice that some of the selection steps in Table~\ref{tab:subs} might bias our compact core abundance estimates. For instance, while the use of a given redshift range is not expected to impact our results, selections based on core elongation and SDSS-image resolution might exclude some genuine compact cores from the number density estimates. Therefore, one should regard abundance estimates based on the compact core sample in Table~\ref{tab:subs} as lower limits. In \S~\ref{sec:numberdens}, we discuss how removing some selection steps listed in Table~\ref{tab:subs} can be used to obtain upper limits on the number densities.

\subsubsection{The Reference Elliptical Galaxies}
\label{sec:ellipt}

Throughout this study, the cores resulting from galaxy decomposition are treated as if they were independent spheroids detached from their surrounding halos or disks. We compare the properties of these compact cores with a reference sample of bona-fide ellipticals.

Bona-fide ellipticals are selected from the subsample of 456,169 galaxies obtained after applying the first three selections listed in Table~\ref{tab:subs} to the Mendel14 catalogue. {\it Reference elliptical galaxies} have been selected according to their morphology and the type of fitted surface-brightness profiles. We require the galaxy morphology to be {\it Elliptical} according to the definitions given in \S~\ref{sec:numberdens} and Tab.~\ref{tab:morphoType}. Additionally, ellipticals must belong to the {\it Type 1}, one of the four classes devised by \cite{Men14} to divide galaxies according to their profile. {\it Type 1} are single-component galaxies, in which the bulge component dominates the surface-brightness profile at all radii (i.e. B/T$\sim$1). The resulting sample contains 18,916 elliptical galaxies. Size and mass parameters for these reference ellipticals are taken from the whole galaxy fit, using single component, free $n$-Sersic models. Note that a small fraction of ellipticals ($\sim$4.5 \%) are found to host a compact core. However, this does not affect our conclusions, our main aim being to compare the properties of ellipticals as a whole with those of compact cores, seen as separate structures, and red nuggets.

\subsection{The red nugget sample}
\label{sec:redsample}
 
\begin{figure}
  \includegraphics[width=\columnwidth]{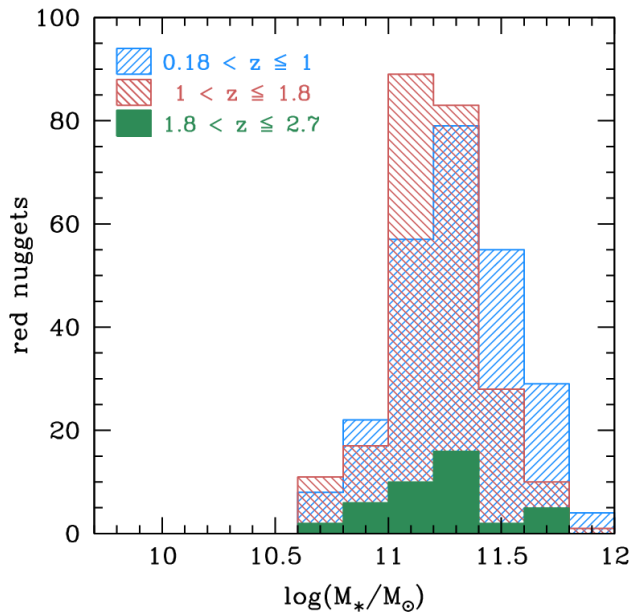}
  \caption{Histogram of the 534 red nuggets that fulfil the vDokkum15 compactness criterion, separated into three redshift bins and shown as a function of their stellar mass.}
  \label{fig:histored}
\end{figure}

Our sample includes observed red nuggets from 29 works in the
literature, 24 of which were part of two major
compilations by \cite{Dam11} and \cite{Sande13}, who renormalized them
to a common observational frame. \cite{Dam11} contributed 433 galaxies with spectroscopic redshifts spanning a range 0.18 $\lesssim$ z $\lesssim$ 2.67. Their compiled stellar masses were re-scaled to a
\citet{BGIMF} IMF. However, the stellar masses of our study
were derived by Mendel14 using M/L ratios for a Chabrier IMF. In order
to carry out a consistent comparison, we have converted \cite{Dam11}
stellar mass values into the Chabrier IMF counterparts. The M/L
conversion follows the prescriptions presented by \citet[Table
  2]{Ber10}, where masses with Chabrier IMF are 0.055 dex heavier than
those estimated with a Baldry \& Glazebrook IMF. Effective radii are
circularized, as in Equation~\ref{eq:radius}, and converted into
physical sizes. The velocity dispersion measurements were extracted
from the original studies and corrected to a common aperture of one
effective radius, using the prescription of \cite{Cap06}. 

Traditionally, red nuggets have been loosely defined as high-redshift (z$\sim$1.5) quiescent galaxies with high masses $M_{\star}\sim 10^{11} {\rm M}_\odot$ yet small effective radii $R_{\rm e} \sim 1$kpc. In order to avoid any ambiguity introduced by a free interpretations of the above definition, we have imposed tight constraints on the sample selection based on the vDokkum15 compactness criterion (Equation \ref{eq:vD15}).  

Table~\ref{tab:rednuggets} summarizes the different studies
contributing to our red-nugget sample. Column 3 shows the number of
objects in each sample, followed by the number of galaxies with measured
velocity dispersion, $\sigma$, in parenthesis. The number of galaxies identified as compact (vDokkum15) are shown in the fourth column. In summary, the final sample contains 534 red nuggets (see Figure \ref{fig:histored}), 218 of them with velocity dispersion measurements.

The less strict Barro13 compactness criterion (Equation \ref{eq:barro13}), with its lower mass limit, would generate a larger sample of 769 red nuggets, but at the cost of contaminating the sample with star-forming galaxies. In a study of star-forming galaxy contamination of assumed quiescent galaxy samples, \cite{Mor13} find that, for z > 0.5 and $\log(M_\star/{\rm M}_\odot) < 10.25$, up to 85\% of the galaxies show high specific Star Formation Rates (sSFR), i.e. $\log({\rm sSFR}/{\rm Gyr}^{-1}) > -2$. The precise percentage depends on the quiescence selection criterion. Conversely, for $\log(M_\star/{\rm M}_\odot) > 10.75$, the contamination decreases to 30 \%. From \cite{Mor13} (Figure 10 and Table 2) we can conclude that any quiescent galaxy sample with a cutoff at log($M_{\star})\geq$10.6 will be significantly cleaner than its analogue with log($M_{\star})\geq$10.0. In summary, unless stated differently (e.g. in the number density comparison, \S\,\ref{sec:numberdens} ), our red nugget sample follows the vDokkum15 compactness criterion.

\begin{figure*}
  \includegraphics[width=\textwidth]{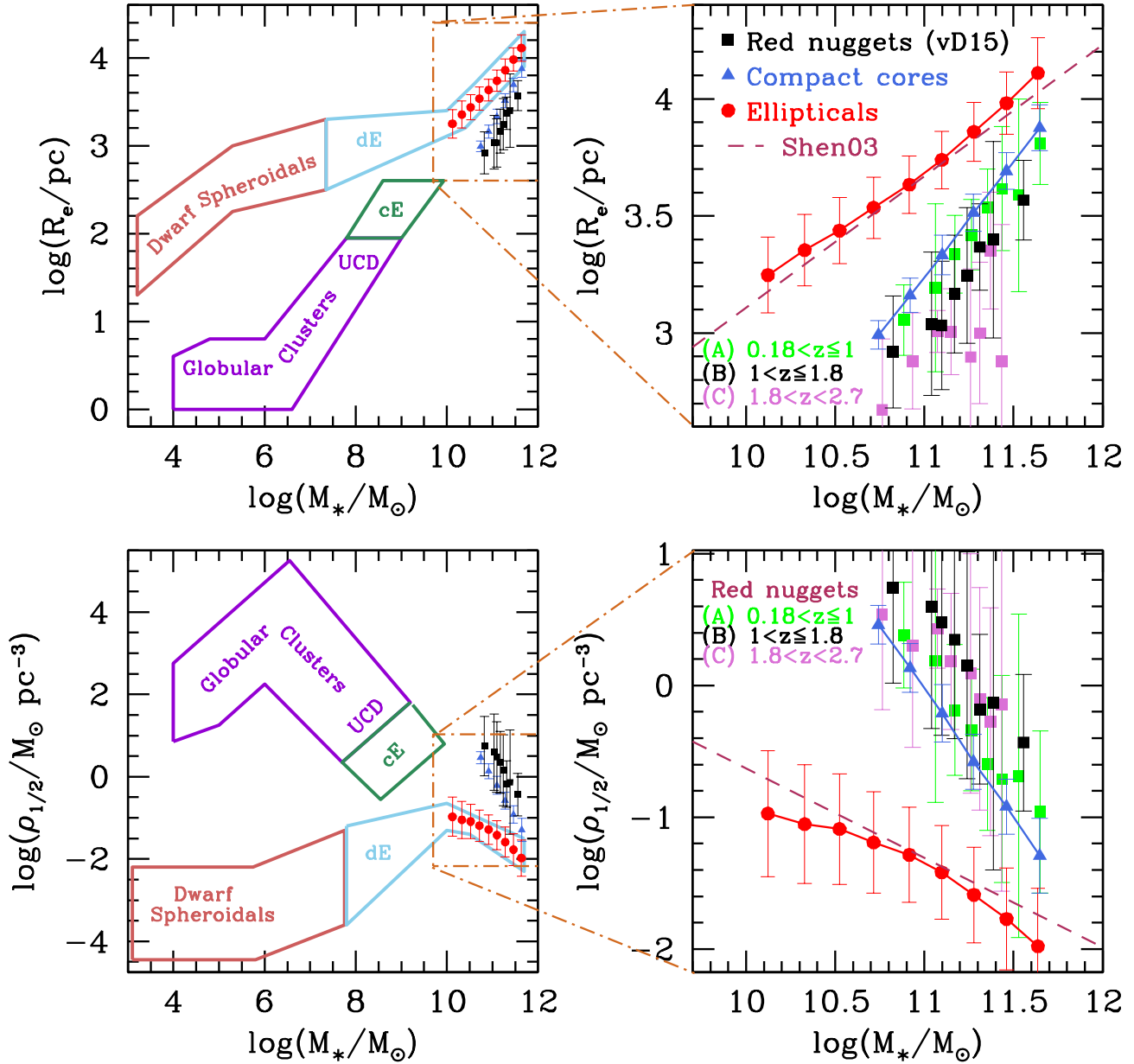}
  \caption{The place of compact cores and red nuggets in the spheroidal family map. The upper-left panel shows the general mass-size plane, including dwarf spheroidals, dwarf ellipticals (dE), normal ellipticals, globular clusters, ultra-compact dwarves (UCD), compact ellipticals (cE). In the upper-right panel, an amplification of our main contribution is shown. Median values of our elliptical galaxies match the results from Shen et al. (2003) and segregate from the compact cores and red nuggets. The red nugget sample has been divided into three redshift slices, whose mass-size relations approach compact cores as redshifts gradually decrease. The lower half of the figure shows the mass-density relation with a similar format.}
  \label{fig:spheroids}
\end{figure*}

\begin{table}
\caption{Data compilation for the red nugget sample. The second column
  shows the redshift range of each contribution, with the number of
  objects displayed in the two last columns. The third column shows the total number of objects from each contribution, followed by the number of objects with measured velocity dispersion, $\sigma$, in parenthesis. With a similar format, the last column shows the number of objects satisfying the vDokkum15 compactness criterion.} 

\label{tab:rednuggets}
\bgroup
\setlength\tabcolsep{\fill}
\begin{tabular}{llrr}
\hline
Contributions & redshift & \multicolumn{2}{c}{Objects}\\
& & Total ($\sigma$) & Compact \\ 
\hline
\multicolumn{4}{c}{\protect\cite{Dam11} Compilation}  \\ 
\hline
\cite{Sag10} & 0.24\ --\ 0.96 & 154 (154) & 70 (70) \\
\cite{Scha99} & 0.29\ --\ 0.99  & 36 (0) & 3 (0) \\ 
\cite{Tre05} & 0.18\ --\ 1.14 & 76 (76) & 30 (30) \\ 
\cite{Ret10} & 1.09\ --\ 1.35 & 44 (0) & 17 (0) \\ 
\cite{New10} & 1.05\ --\ 1.59 & 17 (17) & 7 (7) \\ 
\cite{Mac07} & 1.29\ --\ 1.59 & 5 (0) & 3 (0) \\ 
\cite{Lon07} & 1.22\ --\ 1.70 & 9 (0) & 9 (0) \\ 
\cite{Ryan12} & 1.33\ --\ 1.62 & 6 (0) & 2 (0) \\ 
\cite{Dam11} & 0.62\ --\ 1.75 & 31 (0) & 10 (0) \\ 
\cite{Carr10} & 1.23\ --\ 1.36 & 3 (0) & 3 (0) \\ 
\cite{Dam09} & 1.40\ --\ 1.85 & 10 (0) & 7 (0) \\ 
\cite{Sar11} & 0.96\ --\ 1.91 & 15 (0) & 1 (0) \\ 
\cite{Cas10} & 1.32\ --\ 1.98 & 4 (0) & 1 (0) \\ 
\cite{Cim08} & 1.42\ --\ 1.98 & 8 (0) & 5 (0) \\ 
\cite{Dad05} & 1.39\ --\ 2.67 & 6 (0) & 4 (0) \\ 
\cite{vDo08} & 2.02\ --\ 2.56 & 9 (0) & 9 (0) \\ 
\hline 
\multicolumn{4}{c}{\protect\cite{Sande13} Compilation}  \\ 
\hline
\cite{Sande13} & 1.46\ --\ 2.09 & 5 (5) & 5 (5) \\ 
\cite{Bez13} & 1.24\ --\ 1.62 & 6 (6) & 5 (5) \\ 
\cite{vDo09} & 2.19 & 1 (1) & 1 (1) \\ 
\cite{Ono12} & 1.43\ --\ 1.83 & 18 (1) & 11 (1) \\ 
\cite{Cap09} & 1.42 & 2 (2) & 0 (0) \\ 
\cite{Wel08} & 0.83\ --\ 1.14 & 40 (40) & 14 (14) \\ 
\cite{Tof12} & 1.80\ --\ 2.19 & 1 (1) & 1 (1) \\ 
\hline 
\multicolumn{4}{c}{Others}  \\ 
\hline
\cite{Tru07} & 0.42\ --\ 1.89 & 421 (0) & 227 (0) \\ 
\cite{Kro14} & 1.84\ --\ 2.20 & 14 (0) & 5 (0) \\ 
\cite{Bell14a} & 0.90\ --\ 1.58 & 11 (11) & 11 (11) \\ 
\cite{Bell14b} & 2.09\ --\ 2.43 & 5 (5) & 5 (5) \\ 
\cite{Bez15} & 0.36\ --\ 1.03 & 103 (103) & 2 (2) \\ 
\cite{Zah15} & 0.21\ --\ 0.75 & 150 (150) & 66 (66) \\ 
\hline
 TOTAL & 0.18\ --\ 2.67~~ & {\bf 1,239} (601)~~ & {\bf 534 (218)} \\ 
\hline 

\end{tabular}
\egroup
\end{table}

\section{Results}
\label{sec:results}

\subsection{Photometric Structural Comparison}
\label{sec:photo}

The question we pose here is: are there red nugget-like cores at z$\sim$0.1? 

A considerable number of the z$\sim$0.1 galaxy cores in our sample are found to be structurally similar to the red nuggets. Specifically 10,566 compact cores satisfy the vDokkum15 criterion, and thus share the same mass-size region as the red nuggets, as shown in Figure~\ref{fig:compactness}

In Figure~\ref{fig:spheroids}, built on the spheroidal family portrait by \citet[Figure 1]{Gra13}, we put compact cores and red nuggets in the context of the other known members of the spheroidal family, including globular clusters, ultra-compact dwarves (UCD), compact ellipticals (cE), dwarf spheroidals, dwarf ellipticals (dE) and normal ellipticals (E). The right-hand side plots of Figure~\ref{fig:spheroids} zoom on the selected portrait regions where data from the present study are concentrated.

Mass-size and mass-density relations are presented in Figure~\ref{fig:spheroids}. Stellar masses are computed either for each galaxy as a whole, in the case of ellipticals and red nuggets, or for compact cores as individual structures. The stellar mass density $\rho_{1/2}$ has been defined within the volume containing half each object's light. We take the half-light volume radius to be $r_{1/2}\sim 4/3\times R_{\rm e}$. This relation between the 2D projected effective radius $R_{\rm e}$ and the 3D half-light radius (${\rm r_{1/2}}$) is shown to be valid for most surface brightness profiles and a wide range of Sersic indices \citep[]{Cio99,Wolf10}:
\begin{equation}
\rho_{1/2}\equiv \frac{3 M_\star}{8\pi r_{1/2}^3} \approx \frac{0.05 M_\star}{R_{\rm e}^3}
\end{equation}

For the density computation, we assume that the mass-to-light ratio within the cores is constant with radius, therefore half of the stellar mass of the core is enclosed inside the half-light radius (${\rm r_{1/2}}$). For reference, the Figures include the mass-size and mass-density relations form \cite{Shen03}. Their original relation, $\log(R_{\rm e}/{\rm pc}) = 0.56\log(M_{\star}/{\rm M}_\odot)-2.54$, has been converted from \cite{Kroupa01} IMF to Chabrier IMF, by adding 0.05 to the right-hand side of the equation \citep[Table 2]{Ber10}. As expected, there is a good match between this relation and our reference ellipticals. The slight discrepancy -- a steeper mass-size relation at higher masses and a flatter one for lower masses -- has been explained by \cite{Mos13} in terms of biased size measurements originating from the adopted fitting law (see also Appendix~\ref{app:bias}). Also \cite{LaB10} reported systematically lower effective radii for luminous SDSS galaxies due to sky overestimation.

Median values and RMS dispersion (compact cores and ellipticals) have been computed for stellar mass bins with equal width of 0.2 dex, excluding bins with less than sixty objects. For the red nugget subsample, median values and dispersion were calculated in eight equal size bins. The red nugget sample has been subdivided into three redshift slices, (bin A) 0.18$< z \leq$1; (bin B) 1 $< z \leq$1.8 and (bin C) 1.8$< z <$2.7, the first two with comparable sizes (254 and 239) and the third one with just 41 elements.

Mass-size relations in the upper part of Figure~\ref{fig:spheroids} show that, for galaxies of similar mass, compact cores and red nuggets are 2-5 times smaller than ellipticals, showing also a different slope, $\alpha$ ($R_{\rm e}\propto M_\star^{\alpha}$). On one side, the ellipticals \citep[also][]{Shen03} have $\alpha$=0.56. On the other side, compact cores, bin A and bin B red nuggets have $\alpha$=0.98, 0.96 and 0.95 respectively.
Note that the values of these slopes are a consequence of the selected compactness criterion. For instance, the reported ${\rm \langle \alpha \rangle \sim}$0.96 values are driven by our choice of the vDokkum15 compactness criterion (Equation~\ref{eq:vD15}), where $\alpha$=1. 

The redshift evolution of the intercept in the mass-size relation, i.e. the change in average size at a given stellar mass, has been parameterised by \cite{Tru06b} as $\langle R_{\rm e} \rangle_{M_\star} \propto (1 + z)^{\beta}$. Values of the order of $\beta\sim$ -1.5 have been reported for massive Early Type Galaxies (ETGs) in general \citep{New12,vdW14}. In the frame of our hypothesis, where we compare red nuggets with $z\sim0.1$ compact cores, the mass-size trends of both samples will by definition be very similar, both being subject to the same mass-size selection limits. We therefore find a much smaller effective $\beta\approx-0.7$. For instance, the average size of red nuggets at $\log(M_{\star}/{\rm M}_\odot$) = 11 changes from just 1 to 1.4\,kpc from $ z \sim1.4-0.6$, and the average size of compact cores at this mass at $z\sim0.1$ is just 1.7\,kpc. So the mean size of the sub-population is relatively static, but slowly approaching its selection limit (2\,kpc at this mass; see equation \ref{eq:vD15}). 


The bottom panels in Figure~\ref{fig:spheroids} present the mass-density relation. Again, compact cores and red nuggets, selected with the same vDokkum15 criterion,  are segregated from the reference ellipticals, showing significantly higher densities at a given mass. For instance, at $\log(M_{\star}/{\rm M}_\odot$) = 11, red nugget density $\log(\rho_{1/2}/{\rm M}_\odot {\rm pc}^{-3})$ decreases from 0.56 to 0.16 dex (i.e. 3.6 to 1.4 ${\rm M}_\odot{\rm pc}^{-3}$) from redshift bin B to A. There is an additional decrease to $\log(\rho_{1/2}/{\rm M}_\odot {\rm pc}^{-3})\sim -0.02$dex (i.e. to 0.95 ${\rm M}_\odot {\rm pc}^{-3}$) for z$\sim$0.1 compact cores. This last density is more than 20 times larger than the typical one for elliptical galaxies of similar mass.

We have checked the robustness of our result against two possible
sources of bias, i.e. the presence of AGN that can mimic a compact
core in some galaxies, as well as the choice of a fixed Sersic index
$n_{\rm bulge}$=4. As shown in Appendix~\ref{app:bias}, these do not
affect our results.

In summary, we have shown that a significant fraction of $z\sim$0.1 galaxies contain compact cores structurally similar to red nuggets. The actual value of the fraction depends on the compactness criterion, but we find 5 \% of galaxies contain a core satisfying the most restrictive criterion applied here (vDokkum15).

\subsection{Dynamical Structural Comparison}
\label{sec:dyn}

While stellar masses are dependent on assumptions about the stellar
populations properties (e.g. age, metallicity and IMF), the dynamical masses follow more directly from
observations of velocity dispersion and angular size. In this section, we
will repeat the mass-size and mass-density relations in terms of dynamic mass measurements, comparing compact cores to
red nuggets and elliptical galaxies. Velocity dispersion measurements
in cores are vulnerable to light contamination from the disk/halo
component, leading to artificially low values of $\sigma_{\rm core}$. Our
strategy to avoid core contamination uses stellar population
measurements to single-out disk light entering the spectroscopic
fiber. As shown in Appendix~\ref{app:contamina}, we impose a cutoff at luminosity-weighted ages, ${\rm \langle Age\rangle_L} \geq 9.05$\,Gyr, to select 2,548
uncontaminated compact cores. An aperture correction has been applied
to bring velocity dispersion values to a common system, using
\cite{Cap06} prescription to convert each velocity dispersion into its
equivalent at an aperture of one effective radius. Though
such a prescription may not be accurate for cores, having been
originally derived for luminous spheroids,  the aperture corrections
turn out to be very small, less than $\sim$ 2.2 \% on average, implying that
our results are not significantly affected by the method used to
estimate the aperture corrections. As a further check, we have used an alternative  prescription by \cite{Jor95}, giving nearly equal aperture corrected values, with an insignificant mean offset of 1 km sec$^{-1}$. 

\begin{figure}
  \includegraphics[width=\columnwidth]{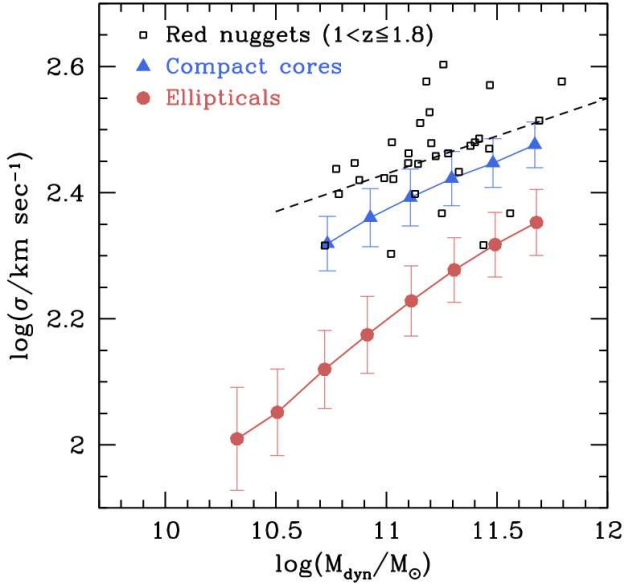}
  \caption{Median values of velocity dispersions ($\sigma$) for
    dynamical mass bins. Samples are compact cores (blue solid triangles),
    and reference elliptical galaxies (red solid dots). Due to the small sample size, the red nuggets in 1$< z \leq$1.8 are represented as individual points (black open squares) with a linear fit shown as a dashed line.}
  \label{fig:veldisp}
\end{figure}

\begin{figure}
  \includegraphics[width=\columnwidth]{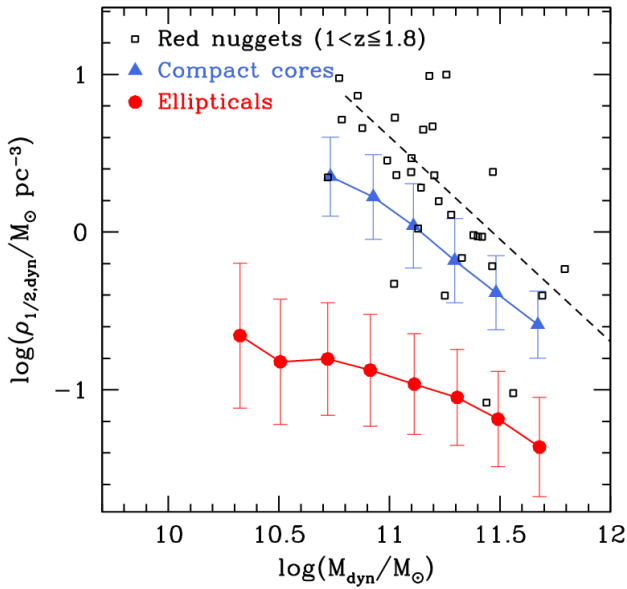}
  \caption{The dynamical version of the photometric mass-density
    relation in Figure~\ref{fig:spheroids} is shown for individual red nuggets
    (black open squares), median values of compact cores (blue solid triangles) and reference elliptical galaxies (red solid dots). The dashed line
    corresponds to the linear fit to the red nuggets. There is a significant  segregation of elliptical galaxies with respect to red nuggets and compact cores .}
  \label{fig:Mdyn}
\end{figure}

The first comparison involves the velocity dispersion ($\sigma$),
the dynamical parameter one might presume to be less affected by the subtleties of
photometric B+D decomposition. However, the measured $\sigma$ are not
exempt from some minor photometric dependencies. Firstly, the
disk-contamination cutoff is based on the stellar mass ($M_{\star}$). Secondly, the $\sigma$ aperture correction is derived from the core angular half-light radius and, thirdly, cores are binned according to their dynamical mass, which depends on the effective radius. Dynamical masses are computed with the formula: 
\begin{equation}
M_{\rm dyn} = K \frac{\sigma^2_{\rm e} R_{\rm e}}{{\rm G}}\label{eq:mdyn}
\end{equation}
with a fixed virial constant $K=5$ deduced by \cite{Cap06} for the ETGs.  

In Figure~\ref{fig:veldisp}, we show the $\sigma$ comparison between three different samples. First, 2,548 compact cores selected according to the vDokkum15 compactness criterion and filtered for disk light contamination, as explained in Appendix~\ref{app:contamina}. Second, the reference elliptical galaxies defined in \S\,\ref{sec:ellipt} also filtered  for light contamination from young stellar populations (8,090 objects) and third, the 28 red nuggets in the redshift slice 1$< z \leq$1.8, with available $\sigma$ observation and $M_{\rm dyn} > M_\star$. From Figure~\ref{fig:veldisp} we can deduce that compact cores and red nuggets are clearly segregated from normal ellipticals. 

The mass-density relation is presented in Figure~\ref{fig:Mdyn} with
median values of compact cores (blue solid triangles) and elliptical galaxies (red solid dots). Again, the small red nugget sample (28) is represented by individual black squares and a corresponding linear fit dashed line. Once again, the relative match between the  compact cores and red nuggets should not come as a surprise, but as a natural consequence of sharing the same compactness criterion. In fact \cite{vD15} show that their compactness criterion corresponds to log$({\rm \sigma_{predicted}/km\  s^{-1}})>$ 2.40.

\subsection{Number Densities}
\label{sec:numberdens}

In this section we compare the abundance of red nuggets at $z
\sim 1.5$ with that of $z\sim 0.1$ compact cores. In \S\,\ref{sec:compl}, we have emphasised the sensitivity of the number density to the choice of
a compactness criterion and the probed mass range, consequently, any number density comparison must be carried out based on given compactness and mass-selection criteria.  We carry out a detailed comparison, by taking advantage of the three available studies on the evolution with redshift of the red nugget abundance \citep[]{Barro13,vdW14,vD15}.

\begin{figure}
  \includegraphics[width=\columnwidth]{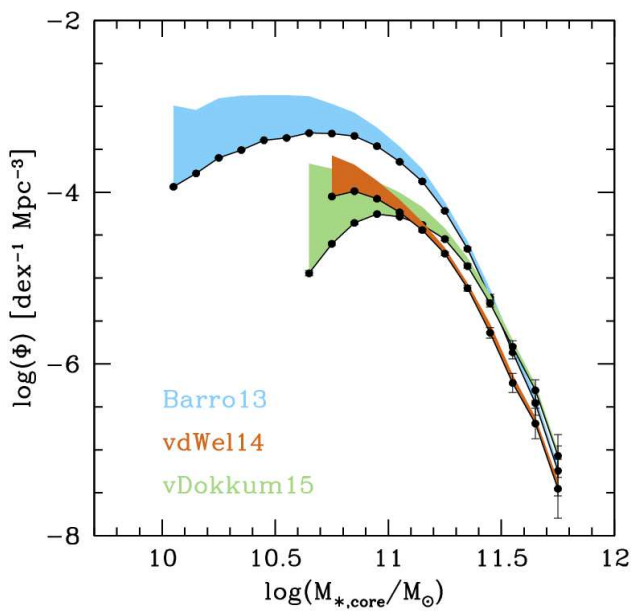}
  \caption{Compact core Mass Function for the three alternative Barro13, vdWel14 and vDokkum15 compactness criteria. Each function spans the range between the upper and lower sample limits, as defined in the text. The lower mass cuts are tied to each compactness criteria definition.}
  \label{fig:MF}
\end{figure}

\begin{table*}
  \caption{Upper and lower number density limits (column 3) for galaxy cores with different compactness criteria and mass ranges. Maximum number densities for red nuggets (column 4) have been extracted from the literature.}
\label{tab:numberdens}
\centering
\begin{tabular}{ | l | c | c | c }
\hline
\multicolumn{2}{c}{Compactness criteria} & \multicolumn{2}{c}{Number density}\\
Reference & Mass range & Compact cores & Red nuggets \\
 &  & low-up limits & maximum \\ \hline
 & $\log(M_{\star,{\rm core}}/{\rm M}_\odot$) & [10$^{-4}\ {\rm Mpc}^{-3}$] & [10$^{-4}\ {\rm Mpc}^{-3}$]\\ \hline
Barro13 & $\geq$ 10.00 & 1.73 - 3.12 & 2.4 $\pm$ 1.0 \\ 
vdWel14 & $\geq$ 10.70 & 0.40 -  0.78 & 1.7 $\pm$ 0.2  \\ 
vDokkum15 & $\geq$ 10.60 & 0.28 - 0.72 & 1.4 $\pm$ 0.1 \\ \hline
\end{tabular}
\end{table*}

We make use of the three compact core samples defined in \S\,\ref{sec:compl} and Table~\ref{tab:subs} in the 8,032\,deg$^2$ area coverage of the SDSS-DR7 Legacy Survey \citep{Aba09}. Compact core Mass Functions (MF) have been computed for each compactness criteria and shown in Figure~\ref{fig:MF}. As stated in Appendix~\ref{app:compl}, V/Vmax volume corrections are applied to account for any residual incompleteness in our samples. Our Vmax calculation follows similar procedures as in \citet{Sim11,Shen03} and Mass Functions are calculated for the redshift range $0.025 \leq z \leq 0.15$. Differences among the three MFs can be understood based on the compactness criteria illustrated in Figure~\ref{fig:compactness}, where one can see, for instance, that the vDokkum15 criteria are more restrictive, at all but at high-masses, than those of Barro13 and vdWel14. As mentioned in \S\,\ref{sec:compl}, the selections in Table~\ref{tab:subs} provide lower limits for the number density. We have worked out corresponding upper limits by reinstating some systems that were excluded when creating the pure sample (due to disk inclination, core elongation, image resolution or Type-4 anomalous decomposition). For instance, filtering by core elongation will exclude genuine compact cores coexisting with bars. The final number of compact cores in the upper limit samples is: 153,038 for Barro13, 25,426 for vdWel14 and 21,830 for vDokkum15 criteria. Figure~\ref{fig:MF} displays the three Mass Functions represented as a band enclosed between the upper and lower limit MFs (black connected dots). 

\begin{figure*}
  \centering
  \includegraphics[width=\textwidth]{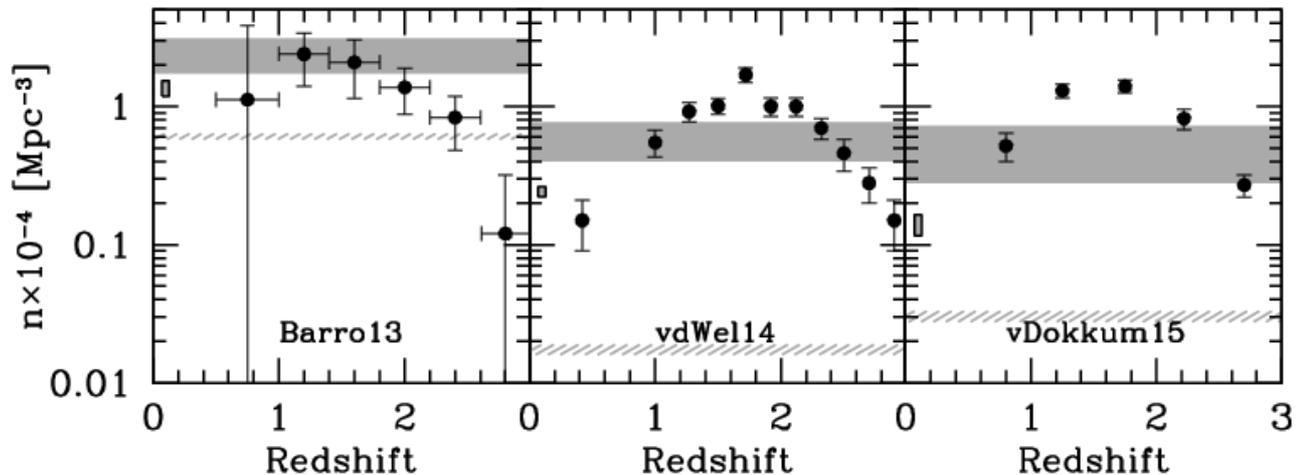}
  \caption{Evolution with redshift of the number density of massive quiescent compact galaxies for the three alternative compactness criteria. The solid dots with error bars are extracted from : Figure 5 of Barro et al. (2013); Figure 13 of van der Wel et al. (2014) and Figure 19 of van Dokkum et al. (2015). Our compact core number density results are represented by a shaded dark grey band confined between the upper and lower limits discussed in the text. Although our result corresponds to redshift $\sim$ 0.1, the band is stretched along the full redshift range to aid the comparison with the high redshift data. The Figure also plots the number density of elliptical galaxies hosting a compact core, shown as a small shaded rectangle at redshift $\sim$ 0.1. Secondly, the number density of entire galaxies, at z $\sim$ 0.1, satisfying the compactness criterion, is shown as a dashed shaded band}
  \label{fig:numberdensity}
\end{figure*}

The lower and upper number density limits can be compared with those for red nuggets, at different redshifts. Solid dots with error bars in Figure~\ref{fig:numberdensity} are extracted form the published plots in \citet[Figure 5]{Barro13}, \citet[Figure 13]{vdW14} and \citet[Figure 19]{vD15}. Dark shaded horizontal bands in Figure~\ref{fig:numberdensity} correspond to the upper and lower limit compact core number density at z$\sim$0.1, stretched along the width of the plot to aid the comparison with the red nuggets. These values correspond to the integral of the upper and lower limits of the mass functions in Figure~\ref{fig:MF}. Results are also reported in Table~\ref{tab:numberdens}. 

Figure~\ref{fig:numberdensity} also plots the number density of the bona-fide ellipticals at z$\sim$0.1 that satisfy, as a whole, the compactness criteria (see dash-shaded region), as well as the number density of $z \sim 0.1$ compact cores residing only in ellipticals (with respect to other galaxy types). The density of compact ellipticals is further discussed in \S\,\ref{sec:discussion}. The low abundance of ellipticals hosting a compact core, with respect to the number density of red nuggets, would refute any proposal that the centre of present-day ellipticals may be the {\bf only} evolutionary path for red nuggets.

From Figure~\ref{fig:numberdensity} we can conclude that the $z\sim 0.1$ compact core number density is comparable to that of red nuggets. In agreement with our results, recent cosmological hydrodynamical simulations on the evolution of red nuggets \citep[]{Wellons15,Furlong15} conclude that only a fraction of them have survived by $z \sim 0$. For instance, \cite{Wellons15} find that: ``about half acquire an ex-situ envelope and exist as the core of a more massive descendant, 30\% survive undisturbed and gain very little mass...''

\subsection{Morphology}
\label{sec:morpho}

The morphological class of the galaxies
hosting compact cores contains information relevant to our study. The \cite{HC11} catalogue provides the
automated morphological classification of the SDSS DR7 spectroscopic
sample, giving the probability P(E), P(S0), P(Sab) and P(Scd) of each
galaxy type. These probabilities can be converted into the T-type
classification scheme of \cite{NaAb10} with the simple linear model
computed by \cite{Meert15},
\begin{equation}
    {\rm Ttype} = -4.6P({\rm E})-2.4P({\rm S0})+2.5P({\rm Sab})+6.1P({\rm Scd})
	\label{eq:Ttype}
\end{equation}
We also follow \citet[Equation 8]{Meert15} in the definition of the galaxy
morphology classes in terms of T-type, which is presented in the
second column of Table~\ref{tab:morphoType}.

\begin{table}
\caption{Morphology class of galaxies hosting a compact core (vDokkum15). The class is assigned according to the definitions in column 2 \citep{Meert15}. Morphology fractions are given for the sample in Table~\ref{tab:subs} (10,566 galaxies) and integrated for $\log(M_{\star,{\rm gal}}/{\rm M}_\odot)\geq 10.50-10.75$ bin.}
\label{tab:morphoType}
\centering
\begin{tabular}{ | c | c | c |}
\hline
Class & Definition &  vDokkum15   \\ \hline
Elliptical & T-type $\leq$ -3 & {\bf 50.9 \%}   \\ 
S0 & -3 $<$ T-type $\leq$ 0.5 & {\bf 43.4 \%}   \\ 
Sab & 0.5 $<$ T-type $\leq$ 4 & {\bf 5.4 \%}   \\ 
Scd & 4 $<$ T-type & {\bf 0.3 \%}  \\ \hline
\end{tabular}
\end{table}

\begin{figure*}
  \centering
  \includegraphics[scale=0.6]{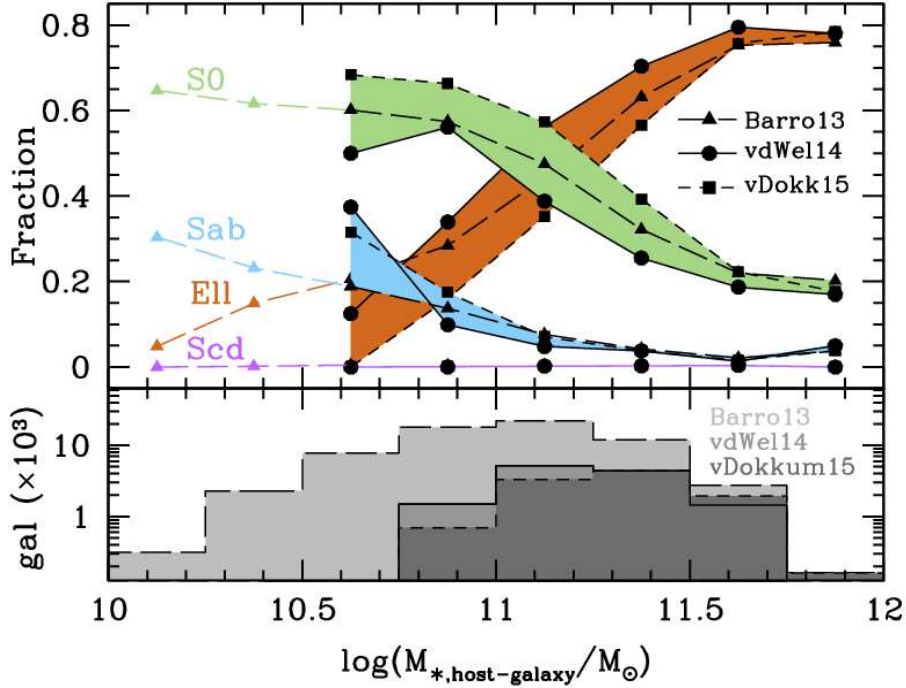}
  \caption{The morphology class fraction of galaxies hosting a compact
core. Both mass and morphology correspond to the whole galaxies. In the upper panel, the results are presented for the three alternative criteria used to select the compact cores: Barro13 (triangles), vdWel14 (dots) and vDokkum15 (squares). Colored bands correspond to each morphology class. The colored triangle extensions correspond to the Barro13 criterion, which includes host galaxies down to $\log(M_{\star, {\rm gal}}/{\rm M}_\odot)\geq$10.0. In the lower panel, three histograms are shown with the total stellar mass of the host galaxy, using a bin size 0.25\,dex. It should be noted that very few host-galaxies populate the 10.5-10.75 mass bin (19 for vDokkum15 and 40 for vdWel14), but nevertheless their morphology fractions have been calculated.}
  \label{fig:morphology}
\end{figure*}

Our goal here is to assess the morphological variety of the fully-fledged galaxies that host compact cores (rather than analyzing the compact core structure itself as in \S\,\ref{sec:photo}). Figure~\ref{fig:morphology} and Table~\ref{tab:morphoType} show these results, with mass estimates, $M_{\star,{\rm gal}}$, also being for the the total host galaxy (not the core mass).

Figure~\ref{fig:morphology} shows the variation along $M_{\star,{\rm gal}}$ of the four morphology class fractions. Aside from the Scd class, which contributes a negligible amount, the other classes are identified by coloured bands. Inside each colour band, lines and symbols identify the results of each compactness criterion. Only the Barro13 results span the mass range down to the $\log(M_{\star,{\rm gal}})\sim 10$ level. The histograms, in the lower part, show the mass range covered by each compactness criterion. Note that very few galaxies, hosting compact cores, populate the $\log(M_{\star,{\rm gal}}/{\rm M}_\odot) = 10.50-10.75$ bin (19 galaxies for vDokkum15 and 40 for vdWel14) and their presence is imperceptible in the histogram. Nevertheless, the morphology fraction is calculated for those galaxies.

Table~\ref{tab:morphoType} gives for morphology fractions integrated across all masses (hence less detailed than the mass-dependant results shown in Figure~\ref{fig:morphology}). Using the vDokkum15 sample in Table~\ref{tab:subs}, all host galaxies with compact cores of log(M$_{\star,{\rm core}})\geq$10.6 (10,566) are integrated into four morphology fraction values. Host galaxy morphology essentially divides into a 50:50 mixture of elliptical and disk morphology (note that the majority of the sample is in early-type systems: E + S0). An alternative option to Figure~\ref{fig:morphology}, using M$_{\star,{\rm core}}$ masses is considered in Appendix~\ref{app:morpho}.

According to Figure~\ref{fig:morphology}, S0 and Sab classes dominate the morphologies of galaxies hosting compact cores, for  log(M$_{\star,{\rm gal}})\leq$11.0. At the high-mass end, elliptical galaxies 
predominate over their disk counterparts. Integrated values in Table~\ref{tab:morphoType} give more balanced results with comparable fractions of Elliptical and S0 classes and minor contributions from later-type disk classes.

A further simple computation can provide an estimation of the average
extra stellar mass incorporated on top of the naked red nuggets to become
fully-fledged present-day galaxies. Assuming that the red nuggets survive
intact since redshift $\sim 1.5$, the extra mass acquired during the
last $\sim$ 9.2 Gyr equates to the difference between the total and
the core masses, roughly equal to the B+D decomposed disk-mass (Md) in
Mendel14.  The average extra mass for the vDokkum15 selected sample is $\log (M_{\star,{\rm extra}}/{\rm M}_\odot)$ = 10.53, which means that an average 25.3$\pm$13.5 \% of the total galaxy mass at z$\sim$0.1 has been accreted since z$\sim$1.5 at a growth rate $\sim 3.7 {\rm M}_{\odot}\,{\rm year}^{-1}$. These results are consistent with those of \cite{vD10}, who estimate that $\sim40\%$ of the total stellar mass at $z = 0$ would have been added through mergers since $z\sim2$ under this hypothesis. There is even better agreement with \cite{Pat13}, who report a 33 \% mass increase from z$\sim$1.5 to z$\sim$0.1.

\section{Discussion}
\label{sec:discussion}

A few recent studies have offered interesting explanations for the early red nugget density evolution at redshifts larger than $z\sim2$. \cite{Barro13} offer a convincing explanation for the number density increase of red nuggets earlier than $z\sim2$, built on the simultaneous decline of the number density of
star forming compact massive galaxies (sCMG). Their result has been corroborated by \cite{vD15}. This interpretation suggests that the sCMG are being continuously converted into red nuggets by processes quenching their star formation, while new sCMGs are replenished by cold flows, until that process becomes inefficient by $z\leq1.5$. Oddly, red nugget evolution below $z\sim1.5$ is characterised by
observational disagreement and the lack of unanimous interpretation.

For instance, our relatively high number density of $z\sim0.1$ compact cores, $(0.28-0.72)\times 10^{-4}\ {\rm Mpc}^{-3}$, (see Table~\ref{tab:numberdens} and Figure~\ref{fig:numberdensity}) come into apparent conflict with the majority of the results from the literature, which range from lower \citep[$\sim 10^{-7}$ Mpc$^{-3}$,][]{Tru09,Tay10} to higher number densities \citep[$\sim 10^{-5}-10^{-4}$ Mpc$^{-3}$,][]{Val10,Pog13}. This conflict is only {\em apparent} because the discrepancy results from the adoption of different search strategies, looking for either compact full galaxies (literature) or compact cores (this study). Contrary to compact cores, which are relatively abundant, very few full galaxies satisfy the demanding compactness criteria in their own right. In order to check this statement we have used our own data to calculate the number density of full galaxies satisfying the vDokkum15 compactness criterion. By using the single-Sersic fit version of the mass-size parameters from the Mendel14 catalogue, we obtain a number density of $3.0\times 10^{-6}\,{\rm Mpc}^{-3}$, an order of magnitude lower than the corresponding compact core number density (as shown in Figure~\ref{fig:numberdensity}). 

According to the hypothesis we are testing here, red nuggets - once formed - do not significantly change their internal structure or composition, but simply aggregate 2D or 3D envelopes to become the compact massive cores of present-day galaxies. Under this hypothesis, the number density of $z\sim0.1$ compact cores should approximately match, or be slightly less than, the maximum value attained by the red nuggets. We have now shown that this is the case (see Figure~\ref{fig:numberdensity} and Table~\ref{tab:numberdens}). This result is in agreement with a recent study by \cite{GraDS15}, their sample of 21 structurally decomposed lenticular galaxies also having a relatively high number density, comparable to our own lower-limits. Comparison with full galaxies at $z\sim0.1$, rather than with compact cores, leads to a discrepancy, as shown here and in the literature. This is in no conflict with the above hypothesis which asserts only that the cores are sufficient -- not necessary -- building blocks for $z\sim 0.1$ galaxies.

\section{Summary and Conclusions}


In this study, we have tested the hypothesis that high-redshift compact, quiescent and massive galaxies, nicknamed {\it red nuggets}, have survived as the compact cores of massive present-day elliptical or disk galaxies. Two main questions are posed to test this hypothesis:

\begin{enumerate}
\item{Do compact cores, structurally similar to red nuggets, exist at $z\sim0.1$?} 
\item{if so, is their abundance comparable to that of red nuggets at $z\sim1.5$?}
\end{enumerate}

We carry out this test using the recently published SDSS-based large catalogues of Bulge+Disk (B+D) galaxy decompositions, in particular from \cite{Men14}, which includes stellar masses of the B and D structural components. We approach the test from a comprehensive perspective, imposing the B+D decomposition to the whole present-day galaxy sample, regardless of their morphology types. We adopt the generic name {\it core} to include the central component of ellipticals and bulge of disks, and study these as if they were independent structures, defined based on their compactness, according to same criteria adopted in the literature to identify the red nuggets themselves \citep[e.g.][]{Barro15,vdW14,vD15}. It is worth mentioning that our results depend heavily on the reliability of the photometric Bulge+Disk decomposition of SDSS galaxy data. Nevertheless, the huge size of the parent catalogue has allowed us to carry out a very strict quality screening, only passed by $\sim\!30$ \% of the objects.  

In summary, the study concludes that the answer to both these questions above is {\em yes}.  The details of this conclusion and other key results are as follows:


$\bullet$ We confirm that a small but significant fraction ($>$5\%) of cores in present-day galaxies (regardless of morphology) is structurally similar to z$\sim$1.5 red nuggets. In the mass-size and mass-density relations, the red nuggets and compact cores are significantly segregated from the normal elliptical galaxies, with average sizes $\sim$4 times smaller and densities $\sim$20 times larger than normal ellipticals at a given mass (e.g. $M_{\star}\sim 10^{11}{\rm M}_\odot$) (Figure~\ref{fig:spheroids}).

$\bullet$ Not only do such compact cores exist, but their abundance matches approximately that of the red nuggets at $z\sim1.5$ \citep[e.g.][]{vD15}. This result is quite consistent with previous findings on the present-day abundance of compact galaxies, as we use of B+D decomposed cores (rather than to full galaxies).

$\bullet$ That these number densities are, if anything, slightly lower than the highest values for red nuggets, further agrees with the hypothesis, leaving room for the some $\sim$20\,\% of red nuggets to have been destroyed in major merger events (as found in some cosmological simulations \citep[e.g.][]{Wellons15,Furlong15}.

$\bullet$ Dynamical measurements of the samples of red nuggets, compact core and ellipticals corroborates the photometric analysis. Velocity dispersion, $\sigma$, and dynamical mass density in red nuggets and compact cores are systematically larger than in normal ellipticals. For instance at a given $M_{\rm dyn} = 10^{11}{\rm M}_\odot$, $\sigma$ is 1.5-1.7 times larger and dynamical-mass density 11-35 times higher than in the normal ellipticals (Figures~\ref{fig:veldisp} and \ref{fig:Mdyn}).

$\bullet$ A simple calculation allows us to estimate the extra mass accreted by the red nuggets to become fully-fledged galaxies, under this hypothesis. Assuming that red nuggets start `naked' or with only rudimentary disks at $z\sim1.5$, the extra stellar mass acquired over the last $\sim\!9$\,Gyr is the difference between the mass of the galaxy and that of the compact core. We find that galaxies hosting a compact core would typically have accreted one quarter of their mass over the last $\sim\!9$\,Gyr, at a growth rate of $\sim\!3.7\ {\rm M}_\odot\ {\rm yr}^{-1}$.
   
$\bullet$ Galaxies hosting compact cores (according to the vDokkum15 criterion) show varied morphological classes extending from ellipticals to spirals (Scd). The integrated fractions of morphology classes with $\log(M_{\rm gal}/{\rm M}_\odot)>$10.5 are: 50.9\% (E), 43.4\% (S0), 5.4\% (Sab) and 0.3\% (Scd). Results for alternative mass ranges and compactness criteria are given in Figure~\ref{fig:morphology} and Appendix~\ref{app:morpho}. These findings are consistent with red nuggets becoming cores of both present-day elliptical and disk galaxies.

\section*{Acknowledgements}

We thank the referee for a very constructive report, which improved the paper. We also thank Ignacio Trujillo, Mar\'\i a Cebri\'an and the Galaxy Group at the IAC for enlightening discussions. We acknowledge the use of SDSS data (http://www.sdss.org/collaboration/credits.html), and support from grants AYA2013-48226-C3-1-P and AYA2013-47742-C04-02-P from the Spanish Ministry of Economy and Competitiveness (MINECO). IF thanks the IAC for hospitality under the Severo Ochoa visitor programme. MS and CdV acknowledge financial support from the MINECO under the 2011 Severo Ochoa Program SEV-2011-0187. CdV, MS and IM-V acknowledge financial support from MINECO under grants AYA2013-46886-P and AYA2014-58308-P.





\begin{thebibliography}{99}
\bibitem[\protect\citeauthoryear{Abazajian et al.}{2009}]{Aba09}
Abazajian, K. N., Adelman-McCarthy, J. K., Ageros, M. A. et al., 2009, ApJS, 182, 543

\bibitem[\protect\citeauthoryear{Baldry \& Glazebrook}{2003}]{BGIMF}
Baldry, I.~K., Glazebrook, K., 2003, ApJ, 593, 258

\bibitem[\protect\citeauthoryear{Baldwin, Phillips \& Terlevich}{1981}]{BPT81}
Baldwin, J. A., Phillips, M. M., Terlevich, R., 1981, PASP, 93, 5

\bibitem[\protect\citeauthoryear{Barro et al.}{2013}]{Barro13} 
Barro, G., Faber, S.M., P\' erez-Gonz\' alez, P.G. et al. 2013, ApJ, 765, 104 

\bibitem[\protect\citeauthoryear{Barro et al.}{2015}]{Barro15} 
Barro, G., Faber, S.M., Dekel, A. et al. 2015, ApJ submitted (arXiv.1503.07164) 

\bibitem[\protect\citeauthoryear{Belli et al.}{2014a}]{Bell14a}
Belli, S., Newman, A. B., Ellis, R. S., Konidaris, N. P., 2014a, ApJ, 788, L29

\bibitem[\protect\citeauthoryear{Belli et al.}{2014b}]{Bell14b}
Belli, S., Newman, A. B., Ellis, R. S., 2014b, ApJ, 783, 117

\bibitem[\protect\citeauthoryear{Berg et al.}{2014}]{Berg14}
Berg, T. A. M., Simard, L., Mendel T. J., Ellison, S. L., 2014, MNRAS, 440, L66

\bibitem[\protect\citeauthoryear{Bernardi et al.}{2010}]{Ber10}
Bernardi, M., Shankar, F., Hyde, J. B., Mei, S., Marulli, F., Sheth, R. K., 2010, MNRAS, 404, 2087

\bibitem[\protect\citeauthoryear{Bezanson et al.}{2009}]{Bez09}
Bezanson, R., van Dokkum, P. G., Tal, T., Marchesini, D., Kriek, M., Franx, M., Coppi, P., 2009, ApJ, 697, 1290

\bibitem[\protect\citeauthoryear{Bezanson et al.}{2013}]{Bez13}
Bezanson, R., van Dokkum, P., van de Sande, J., Franx, M., Kriek, M., 2013, ApJ, 764, L8 

\bibitem[\protect\citeauthoryear{Bezanson et al.}{2015}]{Bez15}
Bezanson, R., Franx, M., van Dokkum, P., 2015, ApJ, 799, 148 

\bibitem[\protect\citeauthoryear{Bluck et al.}{2014}]{Bluck14}
Bluck, A.F.L., Mendel, J.T., Ellison, S.L. et al. 2014, MNRAS, 441, 599 

\bibitem[\protect\citeauthoryear{Buitrago et al.}{2013}]{Bui13}
Buitrago, F., Trujillo, I., Conselice, C. J., H\"aussler, B., 2013, MNRAS, 428, 1460

\bibitem[\protect\citeauthoryear{Cappellari et al.}{2006}]{Cap06}
Cappellari, M., Bacon, R., Bureau, M. et al., 2006, MNRAS, 366, 1126 

\bibitem[\protect\citeauthoryear{Cappellari et al.}{2009}]{Cap09}
Cappellari, M., di Serego Alighieri, S., Cimatti, A. et al. 2009, ApJ, 704, L34

\bibitem[\protect\citeauthoryear{Cardelli, Clayton \& Mathis}{1989}]{CCM89}
Cardelli, J.A., Clayton, G.C., Mathis, J.S., 1989, ApJ, 345, 245 

\bibitem[\protect\citeauthoryear{Carrasco et al.}{2010}]{Carr10}
Carrasco, E. R., Conselice, C. J., Trujillo, I., 2010, MNRAS, 405, 2253 

\bibitem[\protect\citeauthoryear{Cassata et al.}{2010}]{Cas10}
Cassata, P., Giavalisco, M., Guo, Y. et al. 2010, ApJ, 714, L79 

\bibitem[\protect\citeauthoryear{Chabrier}{2003}]{Chab03}
Chabrier, G., 2003, ApJ, 586, L133

\bibitem[\protect\citeauthoryear{Cid Fernandes et al.}{2005}]{Cid05}
Cid Fernandes, R., Mateus, A., Sodr\' e, L., Stasinska, G., Gomes, J.M.,  2005, MNRAS, 358, 363 

\bibitem[\protect\citeauthoryear{Cimatti et al.}{2008}]{Cim08}
Cimatti, A., Cassata, P., Pozzetti, L. et al., 2008, \aap, 482, 21 

\bibitem[\protect\citeauthoryear{Ciotti}{1999}]{Cio99}
Ciotti, L., 1999, \aap, 249, 99 

\bibitem[\protect\citeauthoryear{Conroy, Gunn \& White}{2009}]{Con09}
Conroy, C., Gunn, J. E., White, M., 2009, ApJ, 699, 486

\bibitem[\protect\citeauthoryear{C\^ot\'e et al.}{2006}]{Cote06}
C\^ot\'e, P., Piatek, S., Ferrarese, L. et al., 2006, ApJS, 165, 57

\bibitem[\protect\citeauthoryear{Daddi et al.}{2005}]{Dad05}
Daddi, E., Renzini, A., Pirzkal, N. et al., 2005, ApJ, 626, 680 

\bibitem[\protect\citeauthoryear{Damjanov et al.}{2009}]{Dam09}
Damjanov, I., McCarthy, P.J., Abraham, R.G. et al., 2009, ApJ, 695, 101

\bibitem[\protect\citeauthoryear{Damjanov et al.}{2011}]{Dam11}
Damjanov, I., Abraham, R.G., Glazebrook, K. et al., 2011, ApJ, 739, L44

\bibitem[\protect\citeauthoryear{Damjanov et al.}{2015}]{Dam15}
Damjanov, I., Geller, M.J., Zahid, H.J., Hwang, H.S., 2015, ApJ, 806, 158

\bibitem[\protect\citeauthoryear{Dekel et al.}{2009}]{Dek09}
Dekel, A., Birnboim, Y., Engel, G. et al. 2009, Nature, 457, 451

\bibitem[\protect\citeauthoryear{Di Matteo et al.}{2015}]{Dim15}
Di Matteo, P., G\' omez, A., Haywood, M. et al. 2015, \aap, 577, 1

\bibitem[\protect\citeauthoryear{Dullo \& Graham}{2013}]{DG13}
Dullo, B., Graham, A.W., 2013, ApJ, 768, 36

\bibitem[\protect\citeauthoryear{Fan et al.}{2008}]{Fan08}
Fan, L., Lapi, A., De Zotti, G., Danese, L., 2008, ApJ, 689, L101

\bibitem[\protect\citeauthoryear{Ferreras et al.}{2014}]{Ferr14}
Ferreras, I., Trujillo, I., M\' armol-Queralt\' o, E. et al., 2014, MNRAS, 444, 906 

\bibitem[\protect\citeauthoryear{Furlong et al.}{2015}]{Furlong15}
Furlong, M., Bower, R.G., Crain, R.A. et al., 2015, preprint (arXiv.1510.05645) 

\bibitem[\protect\citeauthoryear{Gadotti}{2008}]{Gad08}
Gadotti, D.A., 2008, MNRAS, 384, 420

\bibitem[\protect\citeauthoryear{Graham}{2013}]{Gra13}
Graham, A.W., 2013, Planets, Stars and Stellar Systems. Vol. 6: Extragalactic Astronomy and Cosmology, ed. T. D. Oswalt \& W. C. Keel (Dordrecht: Springer), 91 

\bibitem[\protect\citeauthoryear{Graham et al.}{2015}]{GraDS15}
Graham, A. W., Dullo, B. T., Savorgnan, G. A. D., 2015, ApJ, 804, 32

\bibitem[\protect\citeauthoryear{Graham \& Worley}{2008}]{GW08}
Graham, A.W., Worley, C.C., 2008, MNRAS, 388, 1708

\bibitem[\protect\citeauthoryear{Hopkins et al.}{2009}]{Hop09}
Hopkins, P. F., Bundy, K., Murray, N., Quataert, E., Lauer, T. R., Ma, C.-P., 2009, MNRAS, 398, 898

\bibitem[\protect\citeauthoryear{Huang et al.}{2013}]{Huang13}
Huang, S., Ho, L. C., Peng, C. Y., Li, Z.-Y., Barth, A. J., 2013, ApJ, 766, 47 

\bibitem[\protect\citeauthoryear{Huertas-Company et al.}{2011}]{HC11}
Huertas-Company, M., Aguerri, J.A.L., Bernardi, M., Mei, S., Sanchez Almeida, J., 2011, \aap, 525, 157 

\bibitem[\protect\citeauthoryear{J\o rgensen, Franx \& Kjaergaard}{1995}]{Jor95}
J\o rgensen, I., Franx, M. \& Kjaergaard, P., 1995, MNRAS, 276, 1341 

\bibitem[\protect\citeauthoryear{Johansson et al.}{2012}]{Joh12} Johansson, P.H., Naab, T., Ostriker, J.P., 2012, ApJ 754 115

\bibitem[\protect\citeauthoryear{Kewley et al.}{2001}]{Kew01} 
Kewley L. J., Dopita M. A., Sutherland R. S., Heisler C. A., Trevena J., 2001, ApJ, 556, 121

\bibitem[\protect\citeauthoryear{Krogager et al.}{2014}]{Kro14}
Krogager, J.-K., Zirm, A. W., Toft, S., Man, A., Brammer, G., 2014, ApJ, 797, 17

\bibitem[\protect\citeauthoryear{Kroupa}{2001}]{Kroupa01}
Kroupa, P., 2001, MNRAS, 322, 231

\bibitem[\protect\citeauthoryear{La Barbera et al.}{2010}]{LaB10}
La Barbera, F., de Carvalho, R.R., de la Rosa, I.G. et al., 2010, MNRAS, 408, 1313

\bibitem[\protect\citeauthoryear{Lackner \& Gunn}{2012}]{LG12}
Lackner, C.N., Gunn, J.E., 2012, MNRAS, 421, L2277

\bibitem[\protect\citeauthoryear{Lauer et al.}{2007}]{Lau07}
Lauer, T.R., Gebhardt, K., Faber, S.M. et al., 2007, ApJ, 664, 226

\bibitem[\protect\citeauthoryear{Licquia \& Newman}{2015}]{LN15}
Licquia, T.C., Newman, J.A., 2015, ApJ, 806, 96

\bibitem[\protect\citeauthoryear{Loeb \& Peebles}{2003}]{LP03}
Loeb, A., Peebles, P. J. E., 2003, ApJ, 589, 29

\bibitem[\protect\citeauthoryear{Longhetti et al.}{2007}]{Lon07}
Longhetti, M., Saracco, P., Severgnini, P. et al., 2007, MNRAS, 374, 614

\bibitem[\protect\citeauthoryear{McGrath et al.}{2007}]{Mac07}
McGrath, E. J., Stockton, A., Canalizo, G., 2007, ApJ, 669, 241 

\bibitem[\protect\citeauthoryear{Meert et al.}{2015}]{Meert15}
Meert, A., Vikram, V., Bernardi, M., 2015, MNRAS, 446, 3943

\bibitem[\protect\citeauthoryear{Mendel et al.}{2014}]{Men14}
Mendel, J. T., Simard, L., Palmer, M., Ellison, S. L., Patton, D. R., 2014, ApJS, 210, 3

\bibitem[\protect\citeauthoryear{Moresco et al.}{2013}]{Mor13}
Moresco, M., Pozzetti, L., Cimatti, A. et al., 2013, \aap, 558, 61

\bibitem[\protect\citeauthoryear{Morishita \& Ichikawa}{2016}]{MorI16}
Morishita, T., Ichikawa, T., 2016, ApJ, 816, 87

\bibitem[\protect\citeauthoryear{Mosleh, Williams \& Franx}{2013}]{Mos13}
Mosleh, M., Williams, R.J., Franx, M., 2013, ApJ, 777, 117

\bibitem[\protect\citeauthoryear{Naab et al.}{2007}]{Naab07}
Naab, T., Johansson, P. H., Ostriker, J. P., Efstathiou, G., 2007, ApJ, 658, 710 

\bibitem[\protect\citeauthoryear{Naab et al.}{2009}]{Naab09}
Naab, T., Johansson, P. H., Ostriker, J., 2009, ApJ, 699, L178

\bibitem[\protect\citeauthoryear{Newman et al.}{2010}]{New10}
Newman, A. B., Ellis, R. S., Treu, T., Bundy, K., 2010, ApJ, 717, L103

\bibitem[\protect\citeauthoryear{Newman et al.}{2012}]{New12}
Newman, A. B., Ellis, R. S., Bundy, K., Treu, T., 2012, ApJ, 746, 162

\bibitem[\protect\citeauthoryear{Nair \& Abraham}{2010}]{NaAb10}
Nair, P.B., Abraham, R. G., 2010, ApJS, 186, 427

\bibitem[\protect\citeauthoryear{Onodera et al.}{2012}]{Ono12}
Onodera, M. et al., 2012, ApJ, 755, 260

\bibitem[\protect\citeauthoryear{Oser et al.}{2010}]{Oser10}
Oser, L., Ostriker, J. P., Naab, T., Johansson, P. H., Burkert, A., 2010, ApJ, 725, 2312

\bibitem[\protect\citeauthoryear{Patel et al.}{2013}]{Pat13}
Patel, S.G., van Dokkum, P.G., Franx, M. et al., 2013, ApJ, 766, 15

\bibitem[\protect\citeauthoryear{Peth et al.}{2015}]{Peth15} 
Peth, M. A., Lotz, J. M., Freeman, P. E. et al., 2015, MNRAS submitted (arXiv.1504.01751)

\bibitem[\protect\citeauthoryear{Poggianti et al.}{2013}]{Pog13}
Poggianti, B. M., Moretti, A., Calvi, R., D'Onofrio M., Valentinuzzi T., Fritz J., Renzini A., 2013, ApJ, 777, 125

\bibitem[\protect\citeauthoryear{Rettura et al.}{2010}]{Ret10}
Rettura, A., Rosati, P., Nonino, M. et al. 2010, ApJ, 709, 512

\bibitem[\protect\citeauthoryear{Ryan et al.}{2012}]{Ryan12}
Ryan, R. E., Jr., McCarthy, P. J., Cohen, S. H. et al., 2012, ApJ, 749, 53

\bibitem[\protect\citeauthoryear{Saglia et al.}{2010}]{Sag10} 
Saglia, R. P., S\' anchez-Bl\' azquez, P., Bender, R. et al. 2010, \aap, 524, A6 

\bibitem[\protect\citeauthoryear{S\' anchez-Bl\' azquez}{2016}]{SB15}
S\' anchez-Bl\' azquez, P., 2016, in Laurikainen E., Peletier R., Gadotti D., eds., Astrophysics and Space Science Library, Vol. 418, Galactic Bulges. Springer International Publishing Switzerland, p. 127 

\bibitem[\protect\citeauthoryear{Saracco et al.}{2011}]{Sar11} 
Saracco, P., Longhetti, M., Gargiulo, A., 2011, MNRAS, 412, 2707

\bibitem[\protect\citeauthoryear{Saulder et al.}{2015}]{Saul15} 
Saulder, C., van den Bosch, R.C.E., Mieske, S., 2015, \aap, 578, 134

\bibitem[\protect\citeauthoryear{Schade et al.}{1999}]{Scha99} 
Schade, D., Lilly, S.J., Crampton, D. et al. 1999, ApJ, 525, 31 

\bibitem[\protect\citeauthoryear{Shen et al.}{2003}]{Shen03} 
Shen, S., Mo, H.J., Simon, D.M. et al. 2003, MNRAS, 343, 978 

\bibitem[\protect\citeauthoryear{Simard et al.}{2011}]{Sim11}
Simard, L., Mendel, J. T., Patton, D. R., Ellison, S. L., McConnachie, A. W., 2011, ApJS, 196, 11

\bibitem[\protect\citeauthoryear{Strauss et al.}{2002}]{Stra02}
Strauss, M. A., Weinberg, D. H., Lupton, R. H. et al., 2002, AJ, 124, 1810

\bibitem[\protect\citeauthoryear{Tacchella et al.}{2015}]{Tacch15}
Tacchella, S. et al., 2015, MNRAS submitted (arXiv:1509.00017) 

\bibitem[\protect\citeauthoryear{Taylor et al.}{2010}]{Tay10}
Taylor, E. N., Franx, M., Glazebrook, K. et al., 2010, ApJ, 720, 723 

\bibitem[\protect\citeauthoryear{Thomas et al.}{2013}]{Thom13}
Thomas, D., Steele, O., Maraston, C. et al., 2013, MNRAS, 431, 1383 

\bibitem[\protect\citeauthoryear{Toft et al.}{2012}]{Tof12}
Toft, S., Gallazzi, A., Zirm, A., Wold, M., Zibetti, S., Grillo, C., Man, A., 2012, ApJ, 754, 3

\bibitem[\protect\citeauthoryear{Treu et al.}{2005}]{Tre05}
Treu T., Ellis R.S., Liao T.X. et al. 2005, ApJ, 633, 174

\bibitem[\protect\citeauthoryear{Trujillo et al.}{2006a}]{Tru06a}
Trujillo, I., Feulner, G., Goranova, Y. et al., 2006a, MNRAS, 373, L36

\bibitem[\protect\citeauthoryear{Trujillo et al.}{2006b}]{Tru06b}
Trujillo, I., F\"orster Schreiber, N.M., Rudnick, G. et al., 2006b, ApJ, 650, 18

\bibitem[\protect\citeauthoryear{Trujillo et al.}{2007}]{Tru07}
Trujillo, I., Conselice, C. J., Bundy, K. et al., 2007, MNRAS, 382, 109 

\bibitem[\protect\citeauthoryear{Trujillo et al.}{2009}]{Tru09}
Trujillo, I., Cenarro, A. J., de Lorenzo-C\' aceres, A. et al., 2009, ApJ, 692, L118 

\bibitem[\protect\citeauthoryear{Trujillo et al.}{2011}]{Tru11}
Trujillo, I., Ferreras, I., de la Rosa, I.G., 2011, MNRAS, 415, 3903 

\bibitem[\protect\citeauthoryear{Valentinuzzi et al.}{2010}]{Val10}
Valentinuzzi, T., Fritz, J., Poggianti, B. M. et al. 2010, ApJ, 712, 226 

\bibitem[\protect\citeauthoryear{van de Sande et al.}{2013}]{Sande13}
van de Sande, J., Kriek, M., Franx, M. et al. 2013, ApJ, 771, 85

\bibitem[\protect\citeauthoryear{van der Wel et al.}{2014}]{vdW14}
van der Wel, A., Franx, M., van Dokkum, P. G. et al. 2014, ApJ, 788, 28

\bibitem[\protect\citeauthoryear{van der Wel et al.}{2008}]{Wel08}
van der Wel, A., Holden, B. P., Zirm, A. W. et al. 2008, ApJ, 688, 48

\bibitem[\protect\citeauthoryear{van Dokkum et al.}{2008}]{vDo08}
van Dokkum, P. G., Franx, M., Kriek, M. et al., 2008, ApJ, 677, L5 

\bibitem[\protect\citeauthoryear{van Dokkum et al.}{2009}]{vDo09}
van Dokkum, P. G., Kriek, M., Franx, M., 2009, Nature, 460, 717 

\bibitem[\protect\citeauthoryear{van Dokkum et al.}{2010}]{vD10}
van Dokkum, P. G., Whitaker, K.E., Brammer, G., et al., 2010, ApJ, 709, 1018 

\bibitem[\protect\citeauthoryear{van Dokkum et al.}{2015}]{vD15}
van Dokkum, P. G., Nelson, E. J., Franx, M. et al., 2015, ApJ, 813, 23

\bibitem[\protect\citeauthoryear{Vazdekis et al.}{2010}]{Vaz10}
Vazdekis, A., S\' anchez-Bl\' azquez, P., Falc\' on-Barroso, J., Cenarro, A. J., Beasley, M. A., Cardiel, N., Gorgas, J., Peletier, R. F., 2010, MNRAS, 404, 1639 

\bibitem[\protect\citeauthoryear{Wellons et al.}{2016}]{Wellons15}
Wellons, S., Torrey, P., Ma, Ch-P. et al., 2016, MNRAS, 456, 1030 

\bibitem[\protect\citeauthoryear{Wolf et al.}{2010}]{Wolf10}
Wolf, J., Martinez, G.D., Bullock, J.S. et al., 2010, MNRAS, 406, 1220 

\bibitem[\protect\citeauthoryear{Zahid et al.}{2015}]{Zah15}
Zahid, H. J., Damjanov, I., Geller, M., Chilingarian, I., 2015, ApJ, 806, 122

\bibitem[\protect\citeauthoryear{Zolotov et al.}{2015}]{Zol15}
Zolotov, A. et al., 2015, MNRAS, 450, 2327

\end{thebibliography}




\appendix

\section{Sample Completeness}
\label{app:compl}

The Mendel14 parent sample spans a redshift range 0.005 $\leq z \leq$
0.4 and covers magnitudes 14.0 $\leq$ mag$_{\rm Petro,r}\leq$ 17.77,
where the bright-end magnitude limit is set to avoid problematic
deblending of luminous galaxies and the faint-end limit marks the
completeness flux-limit of the spectroscopic SDSS sample
\citep{Stra02}. Rather than a flux-limited sample, we need a complete
mass-limited one, given that the most relevant reference parameter of
the present study is the core stellar mass ($M_{\star,{\rm core}}$). 
Although we use the V/Vmax statistics to correct for residual
incompleteness in number density estimations 
(\S\,\ref{sec:numberdens}), it is advisable to rely on a sample with the
highest possible degree of completeness throughout our target mass range,
$\log (M_{\rm core}/{\rm M}_\odot)\geq 10.0$.

\begin{figure}
  \includegraphics[width=\columnwidth]{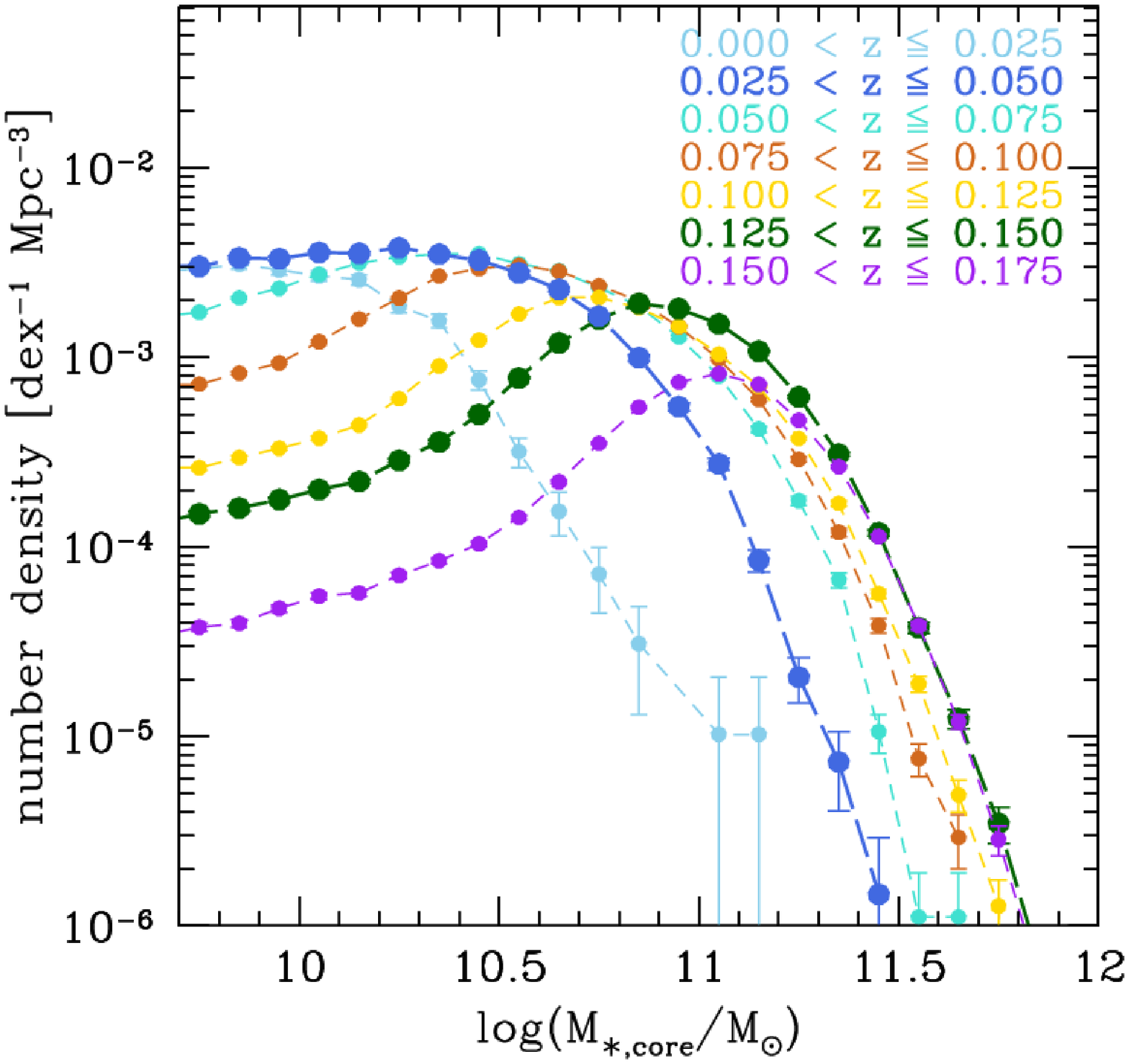}
  \caption{The Core Mass Function (CMF) for redshift slices of
    $\Delta z$ = 0.025 in the parent Mendel14 sample. Our strategy to
    obtain a nearly complete sample for $\log(M_{\rm core}/{\rm M}_\odot)\geq 10.0$ 
    is based on avoiding redshift ranges
    which are poorly populated due to incompleteness. For instance,
    the z $\leqslant$ 0.025 range is excluded due to high incompleteness at
    $\log(M_{\rm core}/{\rm M}_\odot)\geq 10.0$. The highest redshift bin is
    equally excluded and the selected redshift range for the sample is
    0.025$\leqslant z \leqslant$0.150.}
  \label{fig:complete}
\end{figure}

To assess the sample completeness, we have worked out the Core Mass
Function (CMF) for redshift slices of $\Delta z = 0.025$. (see
Figure~\ref{fig:complete}). To construct the CMF we have used the
Mendel14 decomposition parameter {\it logMb}, corresponding to the
logarithm of $M_{\star,{\rm core}}$, instead of the more common
total stellar mass of the galaxy. The area coverage of the SDSS-DR7
Legacy Survey \citep{Aba09} is 8,032\,deg$^2$.

As a first approximation, the upper envelope of all the curves is
taken as our reference CMF. Our strategy to obtain a nearly complete
sample for $\log (M_{\rm core}/{\rm M}_\odot)\geq 10.0$ is based on avoiding
redshift ranges which are poorly populated due to incompleteness. The
redshift range $0.025 \leq z \leq 0.150$ attains the highest
completeness level for the $\log (M_{\rm core}/{\rm M}_\odot)\geq 10.0$ mass-limit. The
limiting CMFs for $0.025 \leq z \leq 0.15$ and $0.125 < z \leq 0.150$
are highlighted in Figure~\ref{fig:complete} to enhance their contrast
with the excluded neighbour redshift slices. The low redshift slice 
$z\leq 0.025$ is rather incomplete at its high-mass end for 
$\log (M_{\rm core}/{\rm M}_\odot)\geq 10.0$, and the high redshift slice 
$0.150 \leq z \leq 0.175$ is also very incomplete at its low-mass end.

\section{Potential Bias Sources}
\label{app:bias}

Two possible bias sources have been checked for their impact on the
photometric B+D decompositions, namely the AGN emission and the choice
of a fixed de Vaucouleurs shape for the compact cores.

\begin{figure}
  \includegraphics[width=\columnwidth]{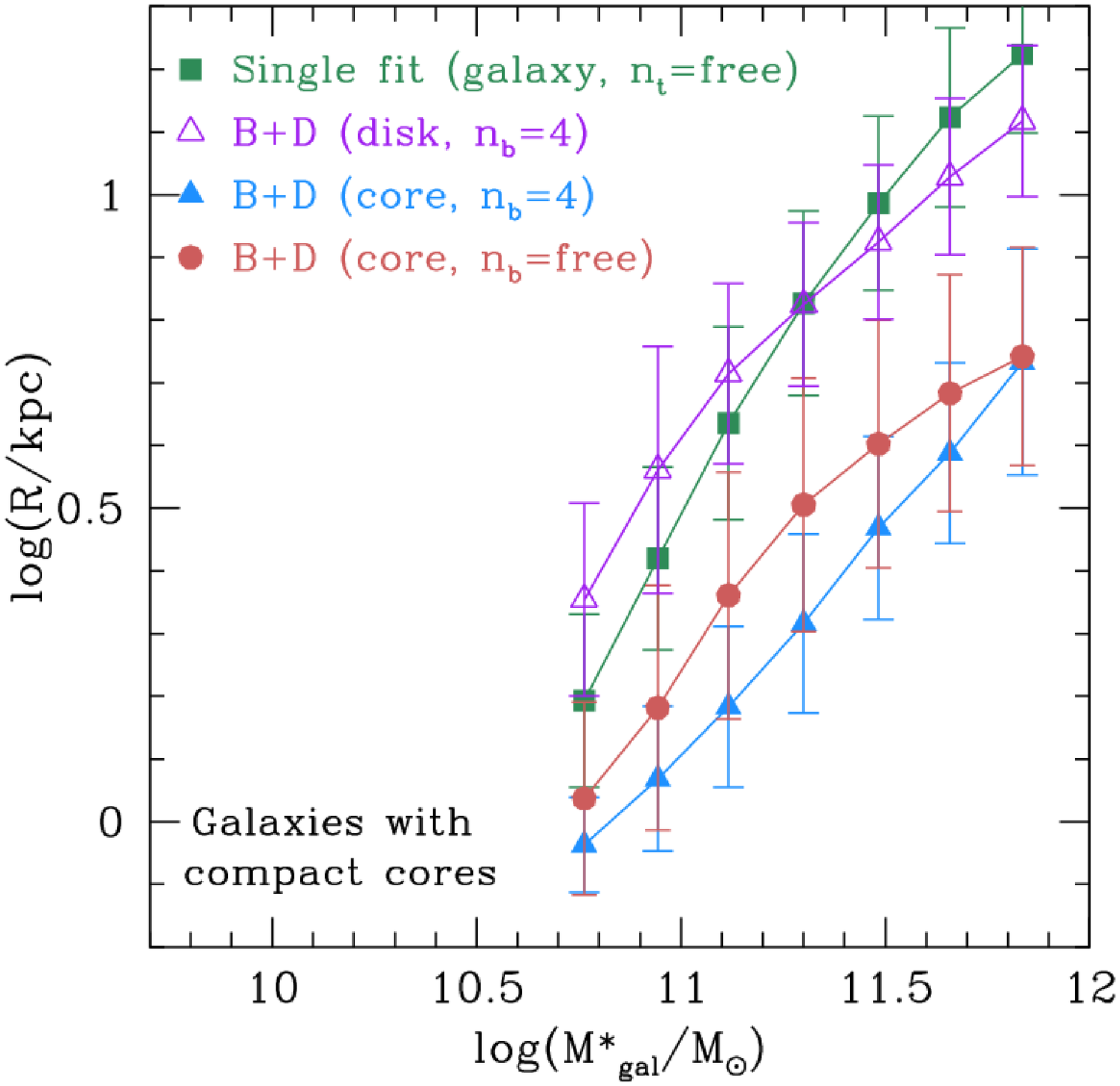}
  \caption{Variation of sizes determined with different fitting laws, for the compact core sample (10,566 objects), plotted as a function of total galaxy mass: Galaxy effective radius measured with a single fit with a free-n Sersic function (green squares); Exponential disk scale length measured with bulge+disk (B+D) decomposition with bulge Sersic $n_b=4$ fit (open purple triangles); core effective radii calculated with $n_b=4$ (blue triangles) and $n_b=$free (red circles).}
  \label{fig:appb_1}
\end{figure}

In particular, the presence of strong emission from active galactic
nuclei (AGN) can bias our measurements of galaxy sizes towards small
values. Following \cite{Thom13}, we have extracted emission line
fluxes to construct the BPT diagram \citep{BPT81}. We
designate an object as {\it emission} when the four diagnostic lines
have been detected with signal-to-noise ratio above 2. The AGN
identification in the BPT diagram follows the empirical separation
from \cite{Kew01} and includes both the Seyfert and LINER
classes. Galaxies designated as {\it emission} (AGN) make the 14 (5)\%
of the compact core subsample, while corresponding figures for the
non-core elliptical subsample are 10(4)\%. The mass-size and
mass-density relations (Figure~\ref{fig:spheroids}), represented by the median values in mass
bins, are unaffected by the exclusion of AGNs or even the exclusion of
{\it emission} objects, proving that AGN contamination is not an issue
for our results.

\begin{figure}
  \includegraphics[width=\columnwidth]{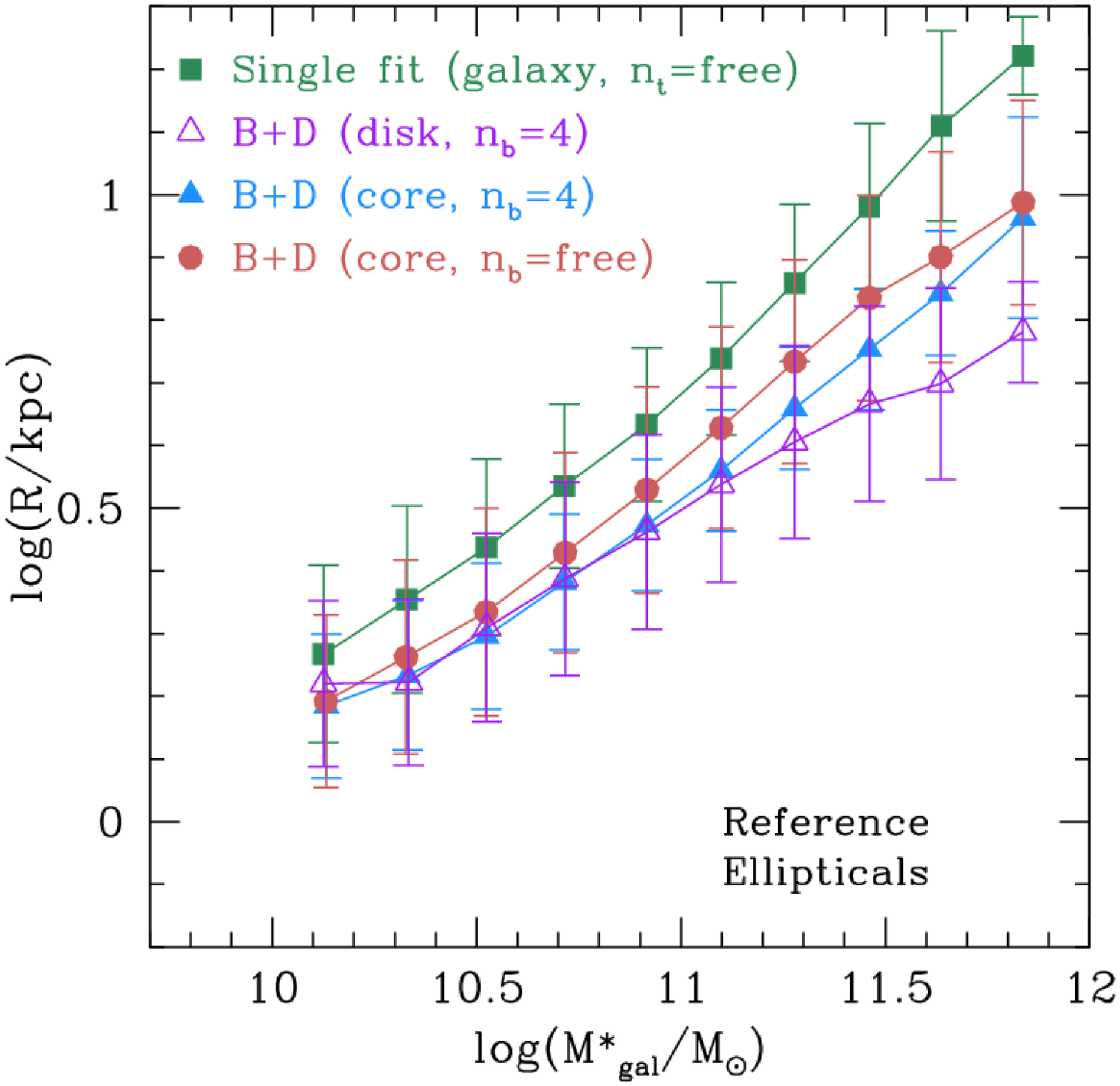}
  \caption{Same as Figure B1, with equivalent symbols, but for the reference elliptical sample.}
  \label{fig:appb_2}
\end{figure}

Concerning the choice of the bulge Sersic index, Mendel14 catalogue
uses a fixed de Vaucouleurs bulge fit ($n_{\rm bulge}$=4), instead of
a free-$n_{\rm bulge}$ fit. According to the authors, the vast
majority of SDSS images have insufficient spatial resolution and/or
signal-to-noise ratio to provide an adequate n-bulge determination and
there is no significant advantage on using free-n$_{\rm bulge}$ over
fixed n$_{\rm bulge}$=4 fits. As explained by \cite{LG12}, at the SDSS
resolution, the bulges have their profile tail subsumed in the disk,
while the central peak is washed out by the PSF, making a bulge with
say $n_{\rm bulge}$=4 indistinguishable from one with 3$\leq n_{\rm bulge}\leq 5$. 

While Mendel14 only derive masses for the $n_{\rm bulge}=4$ B+D fits,
Simard11 provide photometric parameters for both free-n
and fixed n=4 fits. We have used our compact core sample
(10,566 objects) to assess the potential bias introduced by the Sersic
index choice on the structural parameters.

 Alternative galaxy/core size calculations are compared in Figure \ref{fig:appb_1}. As a common base, we have used the total galaxy mass $\log (M_{\star}/{\rm M}_\odot)$ obtained from a single fit with a free-n Sersic function. For $\log (M_{\star,{\rm core}}/{\rm M}_\odot) \geq 10.6$, free-n cores are, on average, $0.11 \pm 0.06$\ dex larger than n=4 cores of similar mass. The use of free-n would decrease the reported separation of $0.38 \pm 0.10$\ dex between the reference ellipticals and compact cores (see Section \ref{fig:spheroids}). As already mentioned, we avoided the use of free-n cores in the present study, due to the lack of $M_{\star,{\rm core}}$ information in the catalogues. In conclusion, the choice of Sersic index does not drastically affect our results. The moderate sensitivity to the Sersic index choice is an expected outcome of the distribution of the Sersic indexes in the free-n fits, which has 78 \% of the values in the interval $3.5\leq$ free-$n_{\rm bulge}\leq 7$, close to the fixed n=4 value. 

In Figure \ref{fig:appb_1}, the behavior of the lines representing the full-galaxy (green) and disk (purple) sizes can be explained in terms of the host galaxy morphology. The slight predominance of disks below $\log (M_{\star}/{\rm M}_\odot) \sim 11.3$ is seemingly related to the larger fraction of disk galaxies in the lower end of our mass interval (see Figure \ref{fig:morphology}). The prevalence of ellipticals at the high mass end generates a drop in the disk sizes.  

\begin{figure}
  \includegraphics[width=\columnwidth]{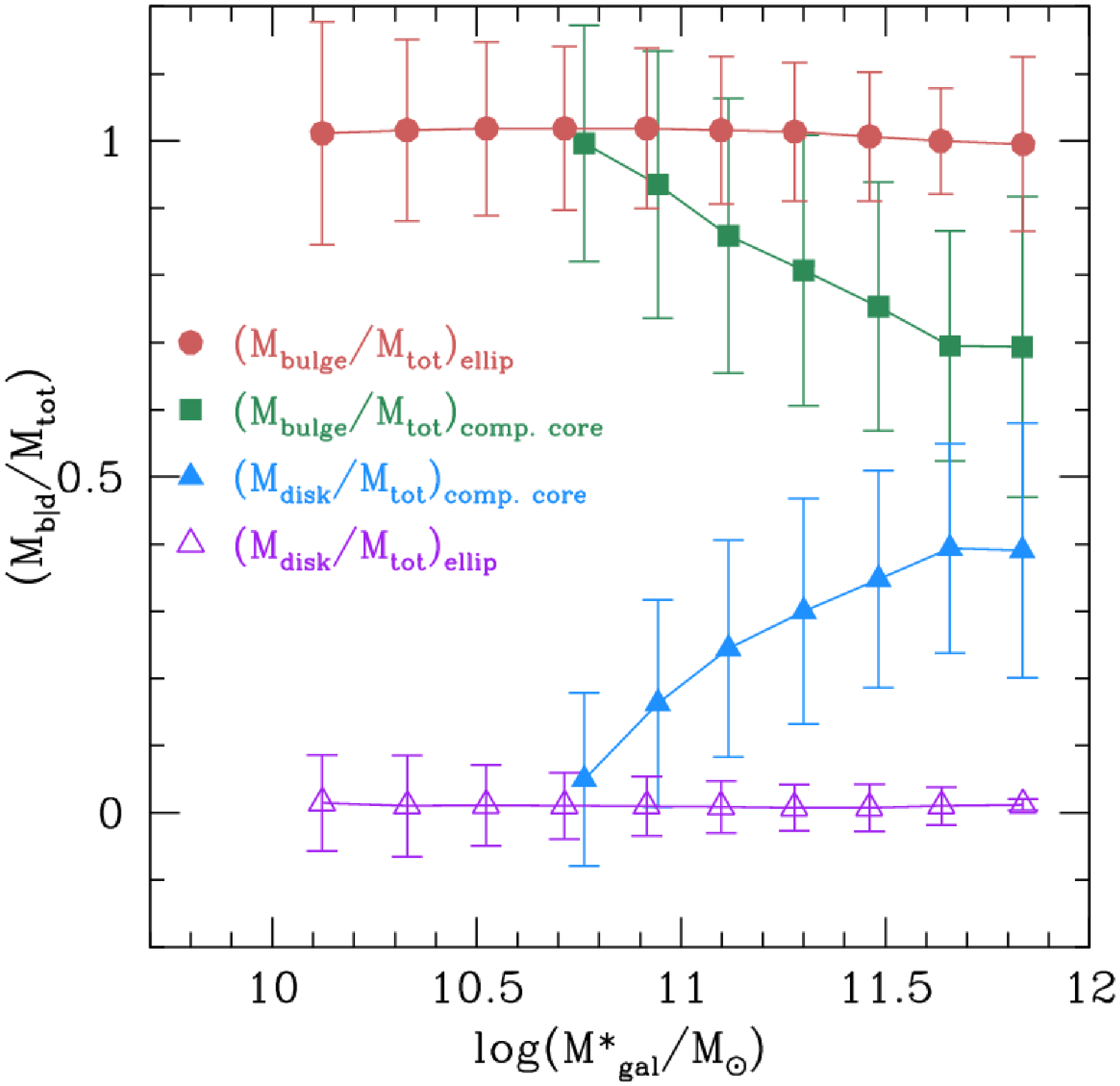}
  \caption{Both compact core and reference elliptical samples, analysed in terms of the mass fraction of their decomposed core (B) and disk (D) structures. A Sersic n=4 fitting law is used. Elliptical galaxies (red circles and open purple triangles) show the expected behaviour of Bulge/Total$\sim$ 1 galaxies. The galaxies hosting compact cores show that the disk/halo accounts for any mass excess above a typical core mass of $\log (M_{\star,{\rm core}}/{\rm M}_\odot) \sim 11$}
  \label{fig:appb_3}
\end{figure}

Figure \ref{fig:appb_2} represents alternative fits to the galaxies of the reference elliptical sample. As expected, the disk size is visibly smaller than the full galaxy and even the cores. Note that the cores of these elliptical galaxies are predominantly extended, only 4.5 \% of them being compact. For the extended cores, the discrepancy between core sizes measured with free-n and n=4 amounts to $0.05 \pm 0.02$\,dex, less than half the offset measured for compact cores (see Figure  \ref{fig:appb_1})

Finally, Figure \ref{fig:appb_3} shows the variation with total mass of the mass fraction contained in the disks and bulges of our two SDSS samples: compact cores and reference ellipticals. As expected from the selection rules, the ellipticals have (B/T) $\sim$ 1 and nearly all their mass is packed in the extended cores. Galaxies hosting a compact core show a distinct behavior in which the more massive the galaxy is, the more relevant the disk/halo becomes. In the frame of our hypothesis, this behavior is compatible with galaxies having a central, prototypical compact core of $M_{\star,{\rm core}}\sim 10^{11}{\rm M}_\odot$, embedded in an accreted disk/halo (whose mass accounts for any excess above the core mass).

\section{Core contamination by disk light}
\label{app:contamina}

In face-on galaxies, contaminant disk light entering the fiber
aperture does not contribute to the velocity dispersion broadening of the core
(i.e. dynamical mass), but contributes to the core stellar mass. Therefore,
disk contaminated cores tend to show an {\it unphysical}
$M_{\star,{\rm core}} > M_{\rm dyn, core}$.
In principle, to minimise disk contamination, we could
restrict $\sigma$ measurements to cores in which the SDSS fiber radius
(1.5 arcsec) were smaller than the core effective radius. This would, however,
be a deficient strategy because the core light-profile frequently gets
submersed below the disk before attaining its half-light radius. For
instance, cores with $R_{\rm e}\lesssim$ 1.5 arcsec could still be
contaminated. Our approach to avoid light contamination is based on
the simple fact that disk light originates in much younger stellar
populations than those from the core \citep[e.g.][and references
therein]{SB15}. Light from a recent star formation event on top of an
essentially old core star formation history (SFH) is an indication of
disk light contamination. Instead of using colours, we adopt the
luminosity-weighted age ${\rm \langle Age\rangle_L}$ as a more
sensitive parameter to detect disk light contamination.

We have carried out a stellar population study in which average ages
(e.g. ${\rm \langle Age\rangle_L}$) and metallicities are obtained
through spectral fitting, using the {\it STARLIGHT} synthesis code
\citep{Cid05} to find the optimal mixture of Simple Stellar
Populations (SSPs) that describes a SDSS spectrum. For the present
study, the basis SSPs correspond to 114 solar-scaled MILES models
\citep{Vaz10}. These models have a Kroupa universal IMF
\citep{Kroupa01} and span a range of six metallicities, from
Z/Z$_{\odot}$= 1/50 to 1.6, and 19 different ages, from 0.5 to 12.6
Gyr. The fitting interval spans from 4000 to 5500 \AA, with emission
lines and bad pixels being masked out. The extinction due to
foreground dust is modelled with the CCM law \citep{CCM89}. Since the SDSS
pipeline does not compute velocity dispersions for spectra with
significant emission lines (as for some disk galaxies), we have also
used the {\it STARLIGHT} spectral fitting to compute our own velocity
dispersion values.

\begin{figure}
  \includegraphics[width=\columnwidth]{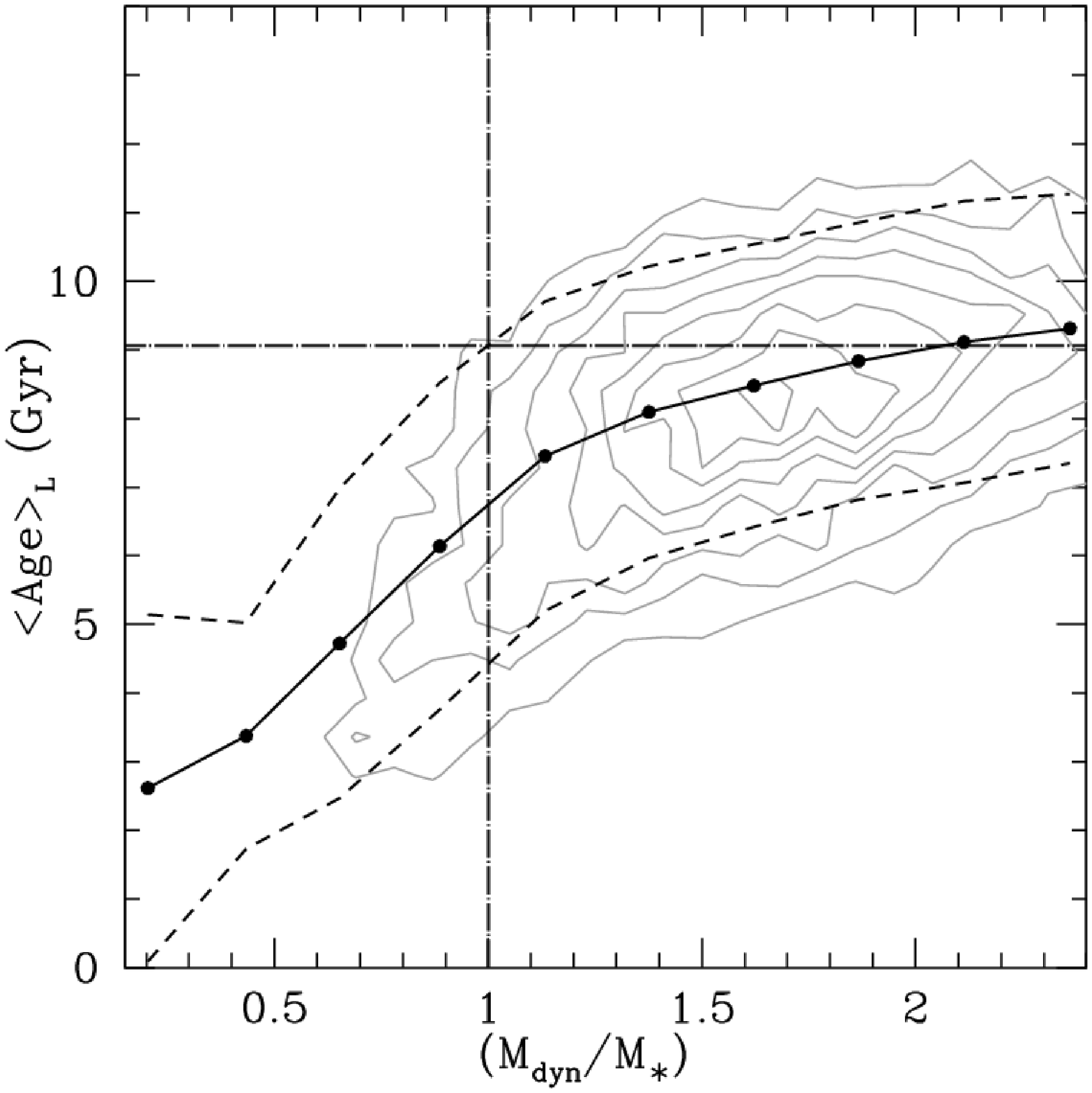}
  \caption{This plot is used to determine a cutoff in ${\rm \langle
      Age\rangle_L}$, by demanding $(M_{\rm dyn}/M_{\star})\geq
    1$. Density contours replace the individual points, while
    connected solid dots represent the median values in $(M_{\rm
      dyn}/M_{\star})$ bins, with dashed lines outlining the
    one-$\sigma$ RMS dispersion around the median. At $(M_{\rm
      dyn}/M_{\star})$ = 1.0, a rather conservative ${\rm \langle
      Age\rangle_L} + 1\sigma$ = 9.05 Gyr cutoff value has been
    obtained.}
  \label{fig:contamina}
\end{figure}

Figure~\ref{fig:contamina} shows the relation between the ${\rm
  \langle Age\rangle_L}$ and the $M_{\rm dyn}/M_{\star}$ ratio for a
mixed sample of our compact cores and reference ellipticals (29,482). Individual points have been replaced by density contours and the connected solid dots
represent the median values in bins of $M_{\rm dyn}/M_{\star}$, with
dashed black lines outlining the one-$\sigma$ RMS dispersion around
the median. The relation supports the idea that unphysical
$M_{\rm dyn}/M_{\star}<1$ values are connected to low ${\rm \langle
  Age\rangle_L}$ values, i.e. disk contamination.  A rather
conservative cut has been worked out to prevent core contamination, by
selecting the ${\rm \langle Age\rangle_L} + 1\sigma = 9.05$\,Gyr at
which the $M_{\rm dyn}/M_{\star}$ ratio becomes larger than unity.

\section{Alternative morphology results}
\label{app:morpho}

\begin{figure*}
  \includegraphics[scale=0.6]{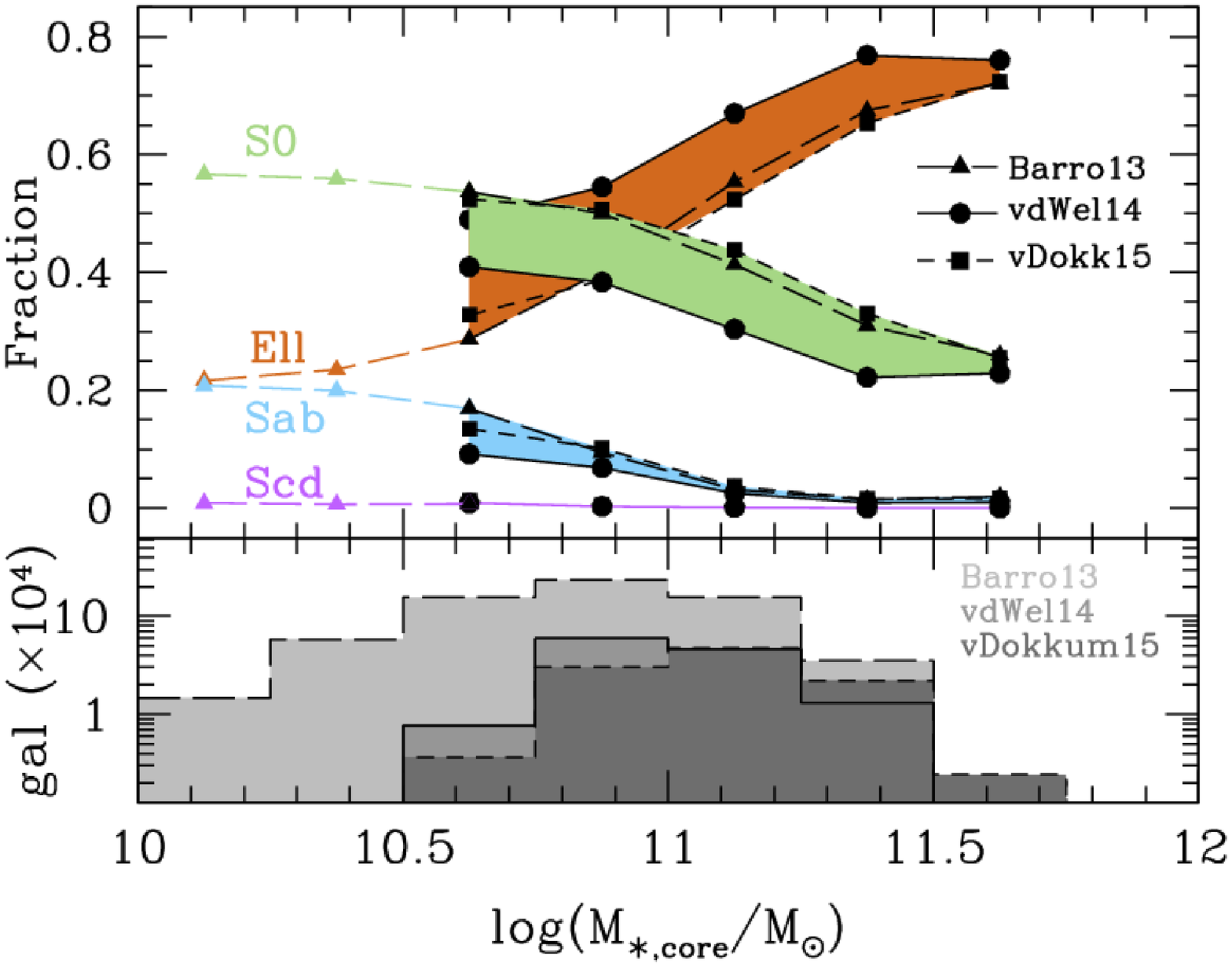}
  \caption{A similar plot to Figure~\ref{fig:morphology}, but using compact core mass $M_{\star,{\rm core}}$ as the x-axis. Morphology refers to the host galaxies while mass corresponds to the compact cores.}
  \label{fig:morpho_Mcore}
\end{figure*}

 Figure~\ref{fig:morphology} displays one of the most important results of the present study. However, an alternative perspective to that result is presented here. using $M_{\star,{\rm core}}$ instead of the $M_{\star,{\rm host}}$. In this case, Figure~\ref{fig:morpho_Mcore}, the plot shape does not change substantially, but it is roughly shifted with respect to those using $M_{\star,{\rm host}}$. The mean $\log(M_{\star,{\rm host}})-\log(M_{\star,{\rm core}})$ is 0.25 dex, so morphology fractions of host galaxies with cores in the $\log(M_{\star,{\rm core}}) = 10.75-11$ bin should be compared with those in the $\log(M_{\star,{\rm host}})=11-11.25$ bin. It is worth emphasising that the integrated results of Figure~~\ref{fig:morpho_Mcore} are identical to those in Table~\ref{tab:morphoType}, because the integration mixes all the galaxies with either $\log(M_{\star,{\rm host}})$ or $\log(M_{\star,{\rm core}}) \geq$10.5.

\bsp	
\label{lastpage}
\end{document}